# Plunging Breakers - Part 1. Analysis of an Ensemble of Wave Profiles

**M. A. Erinin[1,*], X. Liu[1], S. D. Wang[1] & J. H. Duncan[1]†,**

[1]Department of Mechanical Engineering, University of Maryland, College Park, MD 20770, USA

[*]present address: Department of Mechanical and Aerospace Engineering, Princeton University, Princeton, NJ 08544, USA

An experimental study of the dynamics and droplet production in three mechanically generated plunging breaking waves is presented in this two-part paper. In the present paper (Part 1), the dynamics of the three breakers are studied through measurements of the evolution of their free surface profiles during 10 repeated breaking events for each wave. The waves are created from dispersively focused wave packets which are generated by a highly accurate programmable wave maker. The wave maker motions that create the three breakers differ primarily only by small changes in their overall amplitude. Breaker profiles are measured with a cinematic laser induced fluorescence technique covering a streamwise region of approximately one breaker wavelength and over a time of 3.4 breaker periods. The 10 repeated sets of breaker profiles are spatially and temporally aligned to the location and time of jet impact. The aligned profile data is used to create spatio-temporal maps of the ensemble average surface height and the standard deviation of both the local normal distance and the local arc length relative to the instantaneous mean profile. It is found that the mean and standard deviation maps contain strongly correlated localized features and indicate that the transition from laminar to turbulent flow is a highly repeatable process. Regions of high standard deviation include the splash created by the plunging jet impact and subsequent splash impacts at the front of the breaking region as well as the site where the air pocket entrained under the plunging jet at the moment of jet tip impact comes to the surface and pops on the back face of the wave. In Part 2, these features are used to interpret various features of the distributions of droplet number, diameter and velocity.

## 1. Introduction

It is widely accepted that sea spray droplets greatly influence the exchange of mass, momentum, and energy between the ocean and the atmosphere, see for example Melville (1996). In recent years, significant effort has been devoted to studying the sea spray droplet generation through theory, experiments and field measurements, and Veron (2015) and

† Email address for correspondence: duncan@umd.edu

de Leeuw *et al.* (2011) have recently reviewed the subject in detail. Much of the research has focused on the relationship between droplet production and wind conditions. Though it is accepted that droplets are produced primarily by breaking wind waves, only a few studies have focused on the details of the production processes. The present series of two papers is aimed at elucidating a fundamental part of the droplet generation process: the relationship between characteristic events in mechanically generated deep water plunging breaking events and the production and motion of droplets. In this first paper in the series, the breaker behavior is addressed through measurements of the spatio-temporal evolution of the wave profile while the droplet generation is addressed in the second paper with the aid of the material presented herein. In the remainder of this introduction, we focus on previous research on the profile evolution of plunging breakers and the identification of flow structures relevant to droplet production. The introduction to Part 2 will address the literature on droplet production.

A number of papers have examined the profiles of deep-water breaking wave crests in order to explore the incipient breaking conditions, the dynamics of the plunging jet, the series of splash-ups initiated by the plunging jet impact, and the relationship between these quantities and various aspects of the ensuing turbulent flow. A host of parameters have been measured during wave breaking in experiments and simulations including the height of the crest, the depth of the troughs upstream and downstream of the crest, measures of front face and crest-to-trough wave steepness, measures of crest asymmetry, plunging jet trajectories and impact speeds, the height from the wave crest to the plunging jet tip at impact and the area of the air cavity entrapped under the plunging jet at the moment of jet impact. Experimental studies that are directed toward these measurements include Miller (1972), Myrhaug & Kjeldsen (1979), Bonmarin (1989) and Rapp & Melville (1990). Many studies use some of the parameters measured at or leading to the moment of jet impact as independent variables describing the breakers for studies of air entrainment (Lamarre & Melville (1991) and Blenkinsopp & Chaplin (2007)) and dynamic properties of the resulting turbulent flow (Rapp & Melville (1990), Tian *et al.* (2018), Tian *et al.* (2012), Drazen *et al.* (2008) Drazen & Melville (2009) and Lin & Hwung (1992)). Numerical simulations of breaking waves have been quite successful in simulating various aspects of plunging breakers including the turbulent flow, air entrainment and droplet production for relatively short wavelength breakers ($\lambda \approx 30$ to $50$ cm), see Chen *et al.* (1999), Watanabe & Saeki (2002), Watanabe *et al.* (2005), Lubin *et al.* (2006), Derakhti & Kirby (2014), Lubin & Glockner (2015), Pizzo *et al.* (2016), Derakhti & Kirby (2016), Wang *et al.* (2016), Deike *et al.* (2017), and Mostert *et al.* (2022). Some investigators use pre jet impact geometrical parameters to describe the breakers and most present quantities like volume fraction, bubble and vorticity distributions in snapshots at various instants in time. For discussions of these and other studies on plunging breakers, the interested reader is referred to recent review articles by Banner & Peregrine (1993); Kiger & Duncan (2012); Perlin *et al.* (2013). In the discussion of results in the present paper, comparisons with previously published findings will be noted where available.

As mentioned above, the present article is Part 1 of a two-part presentation of the results of an experimental study of droplet generation in plunging breaking waves. In this combined study, spatio-temporal profile and droplet measurements are made in three plunging breakers and the combined results are used for two main goals: 1) to determine the characteristics of the populations of droplets generated in each wave and the correlation of these characteristics with the characteristics of the profile as the crest approaches plunging jet impact and 2) to relate the various times and locations of droplet production and the associated droplet characteristics to events in the breaking process as seen in the profile evolution. In this two-article sequence, the wave profile measurements are discussed herein (Part 1) and droplet measurements are discussed in Part 2. As is explained in Part 2, in order to measure droplets with diameters as small as 100 µm using a cinematic holographic system at various locations in a plane covering about 1.2 breaker wavelengths in streamwise distance, 140 repetitions of each breaker are required. In order to identify events in the profiles that correspond to droplet ejection events localized in time and position, it is necessary to understand the repeatability of the breaking events. To this end, profile measurements over a streamwise distance of 1.1 nominal breaker wavelengths and 2.5 nominal wave periods in time were made with a cinematic LIF system capable of high spatial and temporal resolution. The profile measurements were repeated for 10 realizations of each breaker. These measurements are used to analyze the repeatability of the various events in the breaking process, to track various features in time and position, and to create true ensemble averages of the evolving breaker profile and profile standard deviation. Though many experimental and numerical studies have identified profile phenomena including jet impact, crest phase speed, jet impact speed, and multiple splash up locations, we believe that the examination of the run-to-run repeatability and mean and fluctuating components of the profile is unique to this study. In addition to the above attributes of the work, the profile information is intended to be useful to researchers who may attempt to simulate these experiments with numerical models. Many of these calculations simulate breaking waves in modulated periodic wavetrains rather than the dispersively focusing wave packets used here. For comparisons of, say, droplet production with these numerical studies, the present profile measurements are intended to aid in the adjustment of the breaker generation parameters in the simulations in an effort to create geometrical parameters, like the crest to jet tip height, the jet tip impact speed and the area under the plunging jet, that match the values found in the experiments. In this way, the droplet data can be more effectively compared between the calculations and experiments.

The remainder of this paper is divided into several sections with the experimental details given in § 2, the results and discussion in § 3 and the conclusions in § 4.

## 2. Experimental Details

The data presented in this paper were collected in the wind-wave tank in the Hydrodynamics Laboratory at the University of Maryland. The facilities, experimental methods and data processing techniques used to generated the breaking waves, measure their profile histories and monitor the air-water surface tension are presented in this section. The droplet and humidity measurement techniques are presented in Part 2 of this two-part article.

### 2.1. *Experimental Facility*

The wind-wave tank is 14.8 m long, 1.15 m wide and 2.2 m tall and the experiments were performed with a water depth of 0.91 m, see figure 1. The tank includes a programmable wave maker consisting of a vertically oscillating wedge that spans the width of the tank at one end. The tank also includes an instrument carriage that is supported by hydraulic oil bearings that ride on tracks that are attached to the top of the tank and run parallel to the length of the tank. Most of the optical and camera systems used to measure breaker profiles and droplets in this study are mounted on the instrument carriage. With this mounting system, the optical measurement equipment can be moved without realignment to various locations along the tank length with an accuracy of ± 0.25 mm. During each experimental run, the carriage, and attached measurement systems, remain stationary. A water filtration system consisting of particle filters and a skimmer is used extensively in this experiment to clean the water surface between experimental runs. More details about the experimental facility can be found in Wang *et al.* (2018).

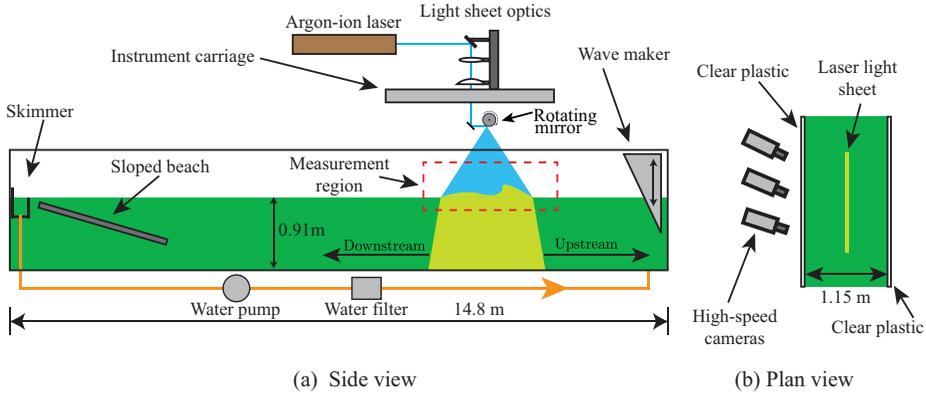

(a) Side view          (b) Plan view

Figure 1: A schematic drawing of side and plan views of the wave tank and LIF wave profile measurement system. See § 2.1 and § 2.2, respectively, for details.

### 2.2. *Breaking Wave Generation*

The wave maker motion used to generate the three breakers is nearly identical to that used in Wang *et al.* (2018) which is based on the dispersively focused wave packet technique first proposed by Longuet-Higgins (1976) and used extensively by Rapp & Melville (1990) and others. In this technique, a packet of linear deep-water gravity waves with varying frequencies is generated such that it converges as it travels along the tank. If the waves are generated with sufficient amplitude, the largest wave crest at the wave packet focal point will break. In the present experiments, the vertical position of the wave maker wedge versus time is given by:

$$z_w = w\,(t)\,\frac{A}{N}\lambda_0 \sum_{i=1}^{N}\left(\frac{k_0}{k_i}\right)^{7/4}\cos\left[x_b\left(\frac{\omega_i}{\bar{c}_g}-k_i\right)-\omega_i t+\frac{\pi}{2}\right],\qquad(2.1)$$

where $w(t)$ is a window function which is described in Wang *et al.* (2018), $A$ is a non-dimensional adjustable constant called the wave maker amplitude, $x_b$ is the nominal streamwise position of the breaking event (as predicted by linear theory) measured from the back of the wedge, $t$ is time, $k_i$ and $\omega_i$ are, respectively, the wavenumber, and frequency of each of the $i = 1$ to $N$ wave components ($\omega_i = 2\pi f_i$ where $f_i$ is the frequency in cycles per second), and $\bar{c}_g$ is the average of the group velocities ($(c_g)_i = 0.5\omega_i/k_i$) of the $N$ components. The frequencies are equally spaced, $\omega_{i+1} = \omega_i + \Delta\omega$, where $\Delta\omega$ is a constant, and the average frequency of the $N$ components is given by $\bar{\omega}$.

For the three breakers studied in the present experiments, $N = 32$, $N\Delta\omega/\bar{\omega} = 0.77$, $f_0 = \bar{\omega}/(2\pi) = 1.15$ Hz, and $\lambda_0 = 2\pi g/\bar{\omega}^2 = 118.06$ cm. The mean position of the wedge is set at $h/\lambda_0 = 0.3579$ (where $h$ is the vertical distance between the still water level and the vertex of the wedge) and the above-mentioned water depth of $H = 0.91$ m corresponds to $H/\lambda_0 = 0.7708$. The variations of the breaking events between the three waves were created primarily by changing $A$. The values of $A$ and $x_b/\lambda_0$ for each of the three breakers are given in Table 1. For values of $A$ slightly less than the minimum value in the table, spilling breakers are produced while for $A$ slightly greater than the maximum value, the first breaker in a given run is located at a distance of approximately one $\lambda_0$ upstream (toward the wave maker) of the main breaker location.

### 2.3. *Breaker Surface Profile Measurements*

Surface profiles of the waves were measured with a cinematic Laser Induced Fluorescence (LIF) technique that has been used extensively in the Hydrodynamics Laboratory, see for example Duncan *et al.* (1999*a*). In this technique, illumination is provided by a 7-Watt Argon-ion laser. The laser beam is focused on the still water surface by two convex spherical lenses and a vertically oriented laser light sheet is created along the center plane of the tank by a 12-sided polygonal mirror rotating at 35,000 rpm. The light sheet is approximately 1 mm thick and 1.6 m long at the still water surface. The water is mixed with a low concentration of Fluorescein dye, which fluoresces with a neon-green color when illuminated in the light sheet.

| $x_b/\lambda_0$ | $A$ | Breaker Intensity |
|---|---|---|
| 6.17 | 0.0651 | Weak |
| 6.20 | 0.0688 | Moderate |
| 6.20 | 0.0707 | Strong |

Table 1: Table of the overall amplitude $A$ and nominal breaking distance $x_b/\lambda_0$ used in the wave maker motion equation (2.1) for the three waves studied herein. All of the other wave maker motion parameters were identical for the three waves, see § 2.2 for details.

Images of the intersection of the laser light sheet with the water surface are captured by three high-speed cameras (Phantom V640 and V641) set up with slightly overlapping fields of view and with time-synchronized image capture. Each camera has a sensor with 2560 × 1600 pixels and 12-bit grey level sensitivity. The combined field of view of the three cameras covers approximately 1.30 m in the horizontal direction and 0.30 m in the vertical direction with a spatial resolution of approximately 180 µm/pixel. This wide field of view covers the plunging jet formation, jet impact and the ensuing turbulent flow region. During each experimental run, a total of 2,000 triplet images were captured at a frame rate of 650 Hz. The resulting recording time is 3.08 seconds, i.e., $3.54T_0$, where $T_0 = 1/f_0$ is the period corresponding to the average wave frequency. The laser beam scans approximately 10 times during the capture of each set of three images. The camera typically has an unobstructed view of the profile formed by the intersection of the light sheet and the water surface including the upper surface of the plunging jet and the jet tip. This profile forms the data presented herein. In this camera view, the profile of the under surface of the jet, including the entrained air cavity, is distorted because it is viewed through the portion of the plunging jet between the plane of the light sheet and the camera. Thus, this region of the water surface profile is not measured. When the water surface is highly three dimensional, the line of sight of the cameras is sometimes blocked by surface structures between the light sheet and the cameras.

To correct for image distortion created by the camera lenses and oblique viewing angles, triplet images of a flat calibration checkerboard, which is placed in the plane of the light sheet, are captured. These calibration images are also used to "stich" the three camera images into a single image with coordinates in the physical plane of the light sheet. It should be noted that the stitching process does not always produce a smooth transition of the surface profile from one camera image to another. These small errors are most noticeable when the free surface becomes rough and 3D structures appear in the image foreground.

The water surface profile in each stitched image triplet is the upper boundary of the wavy bright line at mid height in the images. For the smooth water surface before jet impact, this edge is determined automatically to an accuracy of approximately ±1.0 pixels (±0.2 mm in the plane of the light sheet) for all 650 profiles/s. Using this profile sequence, the time of jet impact is determined to an accuracy of 1/650 s. After jet impact, the regions of the water surface are rough and the determination of the profiles frequently requires operator intervention, a time consuming process. Thus, every eighth stitched image triplet is processed and 81.25 profiles/s are determined. In addition, the start time of the sequence of every eighth processed images was not registered to the moment of jet impact. Thus, the time of each profile relative to the time of jet impact is determined with an accuracy of $1/162.5 = 0.006$ s.

In order to assist those who wish to perform numerical simulations of the three breakers studied herein, measurements of the water surface height versus time at a position 4 m downstream from the back face of the wave maker wedge are provided in Supplementary Material, see file Surface_height_records_at_x=4m.zip.

### 2.4. *Surface Tension Measurements*

It is found that the surface tension as affected by ambient surfactants played a critical role in the breaker behavior. Thus, the surface tension isotherm of the tank water is measured during the experiments to ensure nearly clean-water surface conditions. To accomplish these measurements, samples of the tank water are extracted from below the free surface and placed in a Langmuir trough (KSV NIMA, model KN 1003). The surface tension in the trough is measured with a Wilhelmy plate while the local water surface is slowly compressed by two Teflon barriers that barely touch the water surface and move towards the measurement site at a constant rate. The surface tension before compression was maintained at $73.0 \pm 0.5$ dyn·cm$^{-1}$ (the value for clean water) throughout the experiments, and in all cases the surface tension after compression of the water surface area by 75% over a 60 s period resulted in a drop in surface tension of less than 0.5 dyn·cm$^{-1}$. As the barriers continued to move, creating even higher compressions, the surface tension eventually experienced a sudden drop at compressions ranging from 80% to 95%. The water temperature was also measured and maintained at $\approx 25$ C.

### 2.5. *Experimental Procedure*

The breaker profile experiments reported herein were conducted over a two-day period. Prior to the start of experiments, the tank was filled with filtered tap water mixed with sodium hypochlorite at a concentration of 10 ppm. This high level of chlorination was used to maintain low levels of bacteria and other organic material which are know to produce soluble surfactants. Once filled, the tank water was skimmed and filtered via the diatomaceous filter system for a period of two days. Just before the breaking wave experiments began on the third day, the free chlorine level in the tank was reduced by the addition of hydrogen peroxide ($H_2O_2$) and fluorescein dye was mixed into the tank water at a concentration of approximately 5 ppm. The chlorine concentration reduction is necessary to prevent the chlorine from degrading the fluorescein dye. At the start and end of each day, the surface pressure isotherm was measured using the method described in § 2.4.

Approximately 15 breaking wave events were measured each day. Before each run, the water surface was skimmed for 15 minutes using the diatomaceous earth filtration system while a very light wind in the direction toward the water surface skimmer was applied via the tank's wind tunnel system. After the skimming period, the filter and wind were turned off and the tank water was allowed to come to rest over a 15 minute period. Each experimental run, consisting of the wave maker motion and the imaging of the breaking events, began with a triplet image of the undisturbed water surface.

In order to estimate the amplitude of reflected waves and seiches that might be present during the wave profile and droplet measurements, linear finite-depth wave theory and surface height versus time measurements at a position just upstream of the deep end of the beach were used. It was found that the seiche amplitude was on the order of 0.2 mm and estimated that reflected waves with frequencies of approximately 0.45 Hz and amplitudes less than 0.4 mm would reach the wave profile/droplet measurement region during breaking. Thus, the seiche and reflected waves are insignificant compared the breaking waves whose amplitudes are on the order of 11 cm.

## 3. Results and Discussion

Breaker profile measurements for the weak, moderate, and strong breaking waves are presented and discussed in this section. Ten realizations of the surface profile evolution are measured for each breaker in order to characterize the spatially and temporally evolving mean and fluctuating geometric surface features. In Subsection 3.1, the run-to-run repeatability of the three breakers up to the instant of jet impact is assessed and the alignment of the profile sequences to the profile at the moment of jet impact is described. Then, in Subsection 3.2, the three breakers are characterized by their prominent geometric features during the period from the onset of jet formation to jet impact. In Subsection 3.3, the behavior of the breaker profile after jet impact is presented by examination of the spatio-temporal distribution of the ensemble averaged height and two measures of the profile standard deviation.

The variables used to denote the measured geometric quantities are often accompanied by superscript and subscript notation. The superscripts refer to the time when the quantity was measured while the subscripts describe the profile feature or interest.

### 3.1. *Breaker Repeatability and Profile Sequence Alignment*

The run-to-run repeatability of the breaking wave is assessed using measurements of the wave profile history up to the point of jet impact. After jet impact, the surface profiles exhibit a significant naturally occurring random component which substantially reduces the repeatability and is one of the main issues addressed herein. A sequence of profiles equally spaced in time during jet formation and impact from one realization of the moderate breaker is given in figure 2. The plot details are given in the figure caption. The time of jet formation ($t^f$) is defined as the time when the local tangent to the wave crest profile first becomes vertical at any point on the wave's front face and is determined for each of the ten runs by visually inspecting the breaker profiles. The third profile from the bottom in figure 2 was recorded at a time very close to $t^f$. Similarly, the time of jet impact ($t^i$) is defined as the time when the shape of the tip of the plunging jet changes from convex (when the tip is in the air) to concave on the tip's upper surface shortly after first touching the water surface. The time of jet impact is also determined by visual inspection of the image sequence. The last profile in figure 2 was recorded at a time very close to $t^i$. It is estimated that $t^f$ and $t^i$ are determined to an accuracy of $\pm 1$ one frame in the LIF movies ($\pm 1/650$ s). The values of $t^f$ and $t^i$ relative to their ensemble average values are given for each of the ten runs of the three breakers in table 6 of the appendix. The ranges of $t^f$ and $t^i$ are on the order of seven image frames ($= 7/650$ s) for each of the three waves. In the following, the $(x, y)$ coordinates of the crest point and the jet tip, which are marked by the red triangle and green squares, respectively, on each profile in figure 2, are denoted with subscripts $c$ and $j$, respectively.

Wave crest profiles at $t^f$ and $t^i$ from the ten realizations of the strong breaker are presented in figures 3(*a*) and (*b*), respectively. The zero point of the $x$ coordinate in these plots is taken as the average horizontal position of the point of jet impact. In both plots, it can be seen that the ten wave profiles are quite similar in shape, but the horizontal position of the profiles varies by approximately 3.5 cm which is approximately 3% of $\lambda_0$ and 50% of $\overline{\langle r_x^i \rangle}$, the average horizontal distance from the jet impact point to the crest point, see figure 4. The crest height varies by approximately 1 mm, which is approximately 1.6% of $\overline{\langle r_y^i \rangle}$, the vertical distance from the jet impact point to the crest point. It is thought that these variations are the result of slight variations in wave maker motion and in wave propagation through the tank water, which is probably contaminated by some residual motion from repeated runs. These residual motions include drift currents from the water surface skimming between runs and a small-amplitude seiche motion, which is typical of wave tank experiments with repeated runs, see § 2.5 for more details on the seiche. A table of the standard deviation of the vertical and horizontal positions of the wave crest point at the moment of jet formation, denoted by $x_c^f$ and $y_c^f$, respectively, and the moment of jet impact, denoted by $x_c^i$ and $y_c^i$, respectively, are reported in table 5 of the appendix.

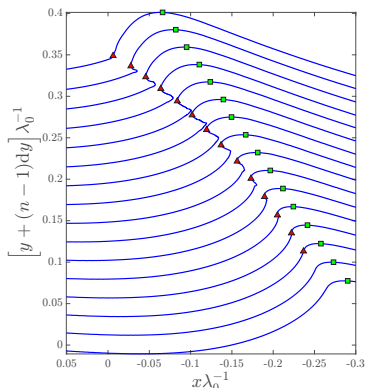

Figure 2: Measured crest profiles from a time shortly before jet formation to the time of jet impact for one realization of the moderate breaker. Each surface profile, shown in blue, is obtained from one LIF image. Note, only the top portion of the plunging jet is visible in the surface profiles as discussed in § 2.3. Each successive profile is separated by a time interval of $\Delta t = 0.0123$ s (every 8th frame in the LIF movie) and plotted $dy = 25$ mm above the previous profile for clarity. The first profile, $n = 1$, is located at the bottom of the plot. The locations of the crest point (the highest point on the wave crest) and jet tip are marked by green squares and red triangles, respectively, on each profile.

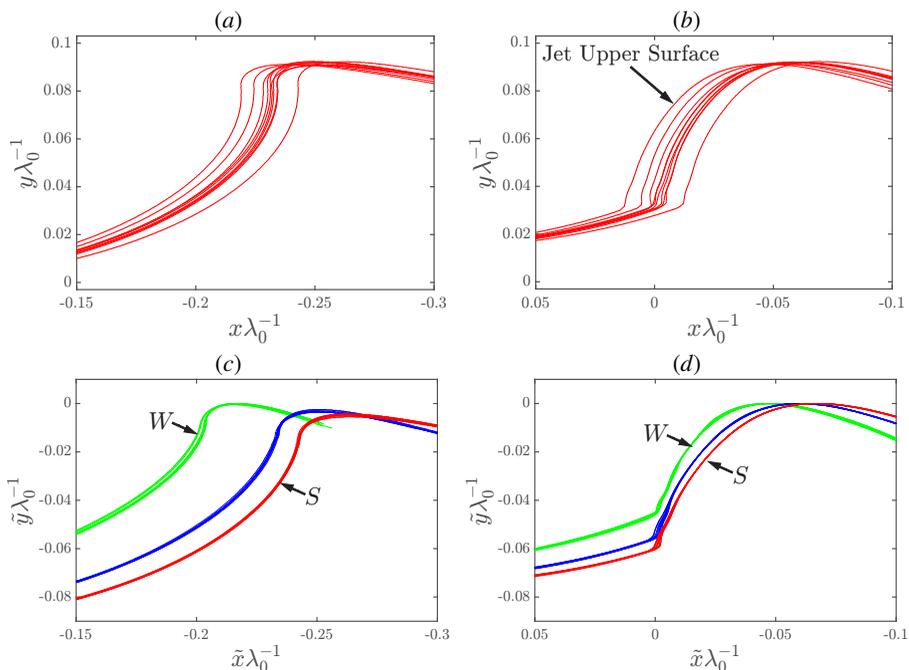

Figure 3: Breaker profiles from ten runs of the strong plunging breaker are shown in red at the time of jet formation and jet impact in subplots ($a$) and ($b$), respectively. The origin of the $x\lambda_0^{-1}$ axis in ($a$) and ($b$) is located at the mean of the 10 streamwise positions of jet impact. The profiles are otherwise not spatially or temporally aligned and therefore give an idea of the run-to-run repeatability of the strong breaker. Spatially and temporally aligned breaker profiles for ten runs of the three breakers are shown in ($c$) and ($d$) for $t = t^f$ and $t^i$, respectively. The weak and strong breakers are indicated by the call-outs $W$ and $S$, respectively. In the axes legends of subplots ($c$) and ($d$), $\tilde{x}\lambda_0^{-1} = (x - \Delta x_b^i)\lambda_0^{-1}$ and $\tilde{y}\lambda_0^{-1} = (y - \Delta y_c^i)\lambda_0^{-1}$. Details of the calculation of the alignment parameters $\Delta x_b^i$, and $\Delta y_c^i$ are discussed in the text.

In order to facilitate the creation of ensemble averaged mean and standard deviation profile histories for each of the three breakers during the ensuing turbulent phase of the breaking events, the profile sequences were aligned in time and space at the moment of jet impact. To perform this alignment, the time in each run is measured relative to the time of jet impact, $\tilde{t} = t - t^i$, and offsets in $x$ and $y$ were determined as follows. The $x$ offsets were determined by sliding the individual profiles at $t^i$ from the 10 runs for each wave horizontally to minimize the difference between each profile and the average profile in the constant slope region (around the mean water level) on the back face of the wave. (This region of the profile was chosen for the alignment because it is nearly a straight line and of highly repeatable slope.) Thus, each profile of height is plotted versus $\tilde{x} = x - \Delta x_b^i$, where $\Delta x_b^i$ is the shift in $x$ required to align each profile with the average profile in the back face region. Recall also that $x = 0$ is the average horizontal position of the jet tip at impact. The vertical coordinate of the aligned profiles is $\tilde{y} = y - \Delta y_c^i$ where the offsets, $\Delta y_c^i$, were determined by moving the profiles vertically, by at most a fraction of 1 mm $= 0.016 \overline{\langle r_y^i \rangle}$, to align all profiles at their crest points with the average crest point height at the moment of jet impact. The variations of positions and times of the crest point and back face from each of the 10 realizations are shown in table 6 in the appendix. The results of this alignment are shown in the plots of the jet formation and jet impact profiles in $(\tilde{x}\lambda_0^{-1}, \tilde{y}\lambda_0^{-1})$ coordinates for the three breakers in figures 3(c) and (d), respectively. As can be seen from the profiles in figure 3(d), the maximum thickness of the band of aligned profiles for each of the three breakers is approximately 3.0 mm $= 0.04 \overline{\langle r_x^i \rangle} = 0.0025\lambda_0$ and occurs in the jet tip region. It is believed that this small region of maximum misalignment is caused by run to run variations in the jet tip shape caused by transverse instabilities in the falling jet as observed in the LIF movies, see Movie 1 given in Supplementary Material, discussed in previous experimental studies including Perlin *et al.* (1996) and analyzed theoretically in Longuet-Higgins (1995). Throughout the remaining regions of the profiles, the band thickness is no more than 1 mm $= 0.014 \overline{\langle r_x^i \rangle} = 0.0009\lambda_0$. This alignment is critical to obtaining a reliable zero level of the standard deviation of the non-breaking part of the breaker profile.

The profiles at $t^f$ are plotted in $\tilde{x}$ - $\tilde{y}$ coordinates in figure 3(c). The 10 crest profiles for each breaker are nearly as well aligned as those at $t^i$. The relative change in the locations of the wave crest in $\tilde{x}$ - $\tilde{y}$ coordinates between jet formation and jet impact indicates that the horizontal distance traveled by the waves increases with increasing breaker intensity as does the increase in crest point height. Details of these results will be given in the following subsection.

### 3.2. *Breaker Characterization up to Jet Impact*

In this subsection, the breaker profile histories are used to obtain quantitative measures of geometric and kinematic parameters describing the three breakers during the time between jet formation and jet impact. Values of many of these parameters at the moment of jet formation and/or jet impact are defined in figure 4 and reported in table 2. This set of parameters is similar to the set defined in figure 3 of Bonmarin (1989) and includes the height from the jet impact point to the wave crest point, herein called $r_y^i$, that was called $h$ and identified in Romero *et al.* (2012), Derakhti & Kirby (2014), Derakhti & Kirby (2016), Deike *et al.*

(2015) and Deike *et al.* (2016) as a key parameter characterizing the post impact breaking wave flows. Several of the measured quantities reported herein will be used later in this paper in correlations with features of the post impact profiles and, in Part 2, with the measurements of droplet production.

Two quantities that describe the large-scale characteristics of the wave profile are the vertical distance $H$ from the lowest point on the trough upstream of the breaking crest, called herein the trough point, to the crest point, and the overall slope of the wave, $H/L$, where $L$ is the horizontal distance from the trough point to the crest point. Values of $\langle H \rangle$, $\langle L \rangle$ and $\langle H/L \rangle$ at $t = t^f$ and $t^i$ are given in Table 2. The wave steepness at jet formation is calculated as $H^f/L^f$, see figure 4. From the values in the table, it can be seen that $H_f$ increases and $L^f$ decreases with increasing breaker strength, resulting in a 16% increase in $H^f/L^f$ from the weakest to the strongest breaker. Comparisons of the measured values of $\langle H^f \rangle$ and $\langle H^f/L^f \rangle$ with the limiting form Stokes wavetrain and previously published experimental data is also useful. For a uniform wavetrain of frequency $f_0$ at the Stokes limit, the wave steepness is $H/\lambda = 0.1411$ (where $H$ is the crest-to-trough height and $\lambda$ is the wavelength) and $g\lambda/(2\pi c^2) = 0.8381$, see for example Longuet Higgins (1984); Cokelet (1977); Schwartz (1974) and Zhong & Liao (2018). Given that $c = f_0\lambda$ and taking $f_0 = 1.15$ Hz, we find $H_{\text{Stokes}} = 0.198$ m, $\lambda_{\text{Stokes}} = 1.193\lambda_0 = 1.408$ m and $c_{\text{Stokes}} = 1.193c_0 = 1.620$ m/s, where $c_0 = g/(2\pi f_0) = 1.358$ m/s. In experimental data from breakers produced by various methods, see Ochi & Tsai (1983); Ramber & Griffin (1987); Bonmarin (1989); Perlin *et al.* (1996), the measured values of $H$ are typically plotted against $gT^2$. In the present experiments, $gT_0^2 = 7.42$ m and at this value the approximate range of $H$ in the above published experiments is from 8 cm to 17 cm. Thus, the values of $H$ in the present experiments (9.8 cm to 10.6 cm) are in the lower range of the values in the literature and all are below the value from the Stokes theory.

The ensemble averaged horizontal and vertical positions of the wave crest point, $\langle \bar{x}_c \rangle \lambda_0^{-1}$ and $\langle \bar{y}_c \rangle \lambda_0^{-1}$, respectively, for the three breakers are plotted versus $\tilde{t}f_0$, in figures 5 (*a*) and (*b*), respectively. The $\langle \bar{x}_c \rangle \lambda_0^{-1}$ versus $\tilde{t}f_0$ data for each of the three breakers form a single nearly straight line. The horizontal speed of the wave crest at jet formation, $\langle u_c^f \rangle$, was calculated by fitting a third order polynomial to the ensemble averaged data set for each wave and evaluating the first derivative of the fitted curves at $t^f$. The values of $\langle u_c^f \rangle$, see Table 2, are nearly the same for the three waves and the average of the three speeds is 1.52 m/s. For reference, consider the phase speeds of linear and limiting form Stokes wave trains, $c_0 = 1.358$ m/s and $c_{\text{Stokes}} = 1.620$ m/s, respectively, as computed in the previous paragraph. The curves of the ensemble averaged crest point height, $\langle \bar{y}_c \rangle \lambda_0$ versus $\tilde{t}f_0$, see figure 5(*b*), have an overall maximum at a time between $\langle t^f \rangle$ and $\langle t^i \rangle$. Near the time of jet formation, $y_c$ increases nearly linearly with time. A third order polynomial was fitted to this ensemble averaged data, see table 7, and the first derivative of the fits were evaluated at the moment of jet formation to yield the rate of rise of the crest, $\langle v_c^f \rangle$, with values of 0.105 m/s, 0.144 m/s, and 0.152 m/s for the weak, moderate, and strong breakers, respectively. From the moment of jet formation to the time when the crest reaches its maximum height, the vertical displacement of the crest point is approximately 8 mm for all three waves. After this maximum, the crest point drops about 2 mm before the moment of jet impact and continues to fall for a short time thereafter, eventually exhibiting irregular motion. The small standard deviation found in this region of the plots indicates the high repeatability of this irregular motion of the crest point.

In the moments leading up to jet formation and impact, the wave becomes asymmetric as its forward face steepens. Just after the moment of jet formation, the jet tip begins to form and moves out ahead of the crest. The jet tip then, as is well known, simultaneously moves forward and falls to the water surface on the front face of the wave, entraining a pocket

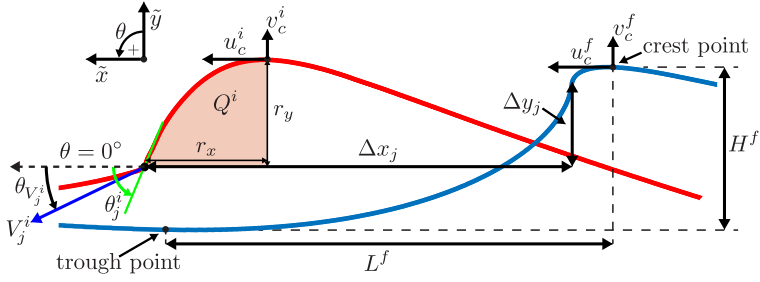

Figure 4: A sketch showing the definition of various geometric and kinematic parameters of the wave profiles at the moment of jet formation, blue profile, and the moment of jet impact, red profile. Numerical values of the parameters for each of the three waves are given in Table 2. The $x$ and $y$ component of the wave crest speed are represented by $u_c$ and $v_c$, respectively, while $V_j^i$ and $\theta_{V_j^i}$ are the speed and angle of the jet tip velocity at impact. The slope of the front face of the jet around the point of jet impact is $\theta_j^i$. The area under the plunging jet at impact, denoted by the red background and labeled $Q^i$, is defined along with the major and minor axes area, $r_x^i$ and $r_y^i$.

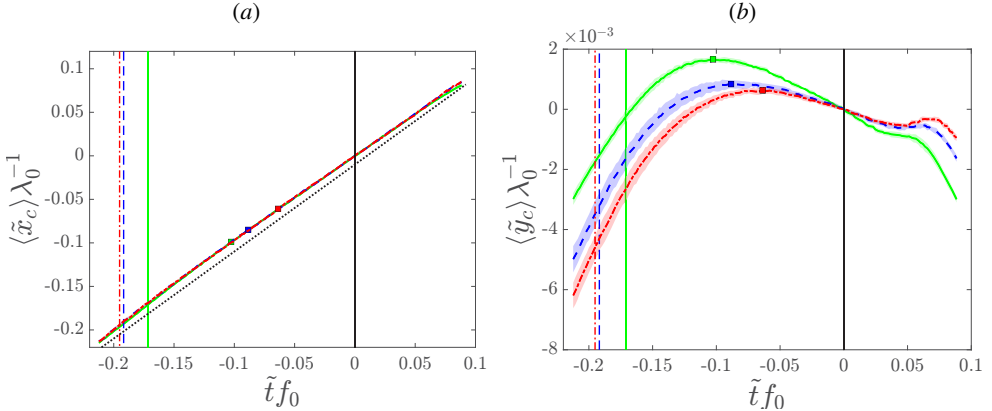

Figure 5: The ensemble average dimensionless horizontal and vertical positions of the crest point, $\langle \tilde{x}_c \rangle \lambda_0^{-1}$ and $\langle \tilde{y}_c \rangle \lambda_0^{-1}$, respectively, are plotted versus $\tilde{t} f_0$ in ($a$) and ($b$), respectively. Data is shown for the weak (green solid line), moderate (blue dashed line) and strong (red dotted line) breakers. The three colored vertical lines indicate the average values of $\tilde{t}$ at the moment of jet formation for each of the three waves, while the vertical black line is the moment of jet impact ($\tilde{t} = 0$) for all three waves. The vertical half width of the color bands on each curve represents the standard deviation of the data at each time instant. The colored bands representing the standard deviation for the data in ($a$) are too small to see in the plot. The slope of the dotted straight line in ($a$) is equal to the phase speed of a linear wave with frequency $f_0$, i.e., $c_p = 1.357$ m/s. The colored squares mark the time of maximum wave crest height on each curve.

of air upon impact. The motion of the jet tip from jet formation to impact is presented in a plot of the ensemble average jet tip height, $\langle \tilde{y}_j \rangle \lambda_0^{-1}$, versus $\tilde{t} f_0$ and a plot of the jet tip trajectory, $\langle \tilde{y}_j \rangle \lambda_0^{-1}$ versus $\langle \tilde{x}_j \rangle \lambda_0^{-1}$, in figures 6($a$) and ($b$), respectively. Similar data for a single realization of a plunging breaker can be found in Drazen *et al.* (2008). A cyan colored ballistic curve is included in figure 6($a$) for comparison with the jet tip data. Third order polynomials were fitted to the ensemble averaged $\langle \tilde{x}_j \rangle \lambda_0^{-1} - \tilde{t} f_0$ and $\langle \tilde{y}_j \rangle \lambda_0^{-1} - \tilde{t} f_0$ points from the time of maximum jet tip height to immediately before jet impact, see table 7. The jet tip velocity components at the moment of jet impact were obtained as the derivatives of these

| Variable | Unit | Weak | Moderate | Strong |
|---|---|---|---|---|
| Jet Formation | | | | |
| $\langle H^f \rangle$ | (mm) | $98.6 \pm 0.3$ | $104.4 \pm 0.3$ | $105.8 \pm 0.4$ |
| $\langle L^f \rangle$ | (mm) | $594.3 \pm 6.5$ | $563.4 \pm 6.7$ | $551.6 \pm 5.1$ |
| $\langle H^f/L^f \rangle$ | | $0.166 \pm 0.0019$ | $0.185 \pm 0.0024$ | $0.193 \pm 0.0022$ |
| $\langle u_c^f \rangle$ | (m/s) | 1.536 | 1.515 | 1.511 |
| $\langle v_c^f \rangle$ | (m/s) | 0.105 | 0.144 | 0.152 |
| Jet Impact | | | | |
| $\langle H^i \rangle$ | (mm) | $84.1 \pm 0.3$ | $89.4 \pm 0.2$ | $90.7 \pm 0.4$ |
| $\langle L^i \rangle$ | (mm) | $569.1 \pm 8.5$ | $548.6 \pm 7.7$ | $529.5 \pm 10.4$ |
| $\langle H^i/L^i \rangle$ | | $0.148 \pm 0.0022$ | $0.163 \pm 0.0022$ | $0.171 \pm 0.0035$ |
| $\langle u_c^i \rangle$ | (m/s) | 1.312 | 1.314 | 1.317 |
| $\langle v_c^i \rangle$ | (m/s) | -0.031 | -0.020 | -0.022 |
| $\langle V_j^i \rangle$ | (m/s) | 1.904 | 1.983 | 2.010 |
| $\langle \theta_{V_j^i} \rangle$ | (degrees) | 21.9 | 23.7 | 27.6 |
| $\langle \theta_j^i \rangle$ | (degrees) | $65.3 \pm 1.2$ | $67.6 \pm 3.3$ | $66.3 \pm 2.6$ |
| $\langle r_x^i \rangle$ | (mm) | $57.4 \pm 2.6$ | $73.5 \pm 2.8$ | $81.5 \pm 1.8$ |
| $\langle r_y^i \rangle$ | (mm) | $53.0 \pm 1.0$ | $65.0 \pm 1.5$ | $69.4 \pm 1.1$ |
| $\langle Q^i \rangle$ | (mm$^2$) | $2222 \pm 140$ | $3457 \pm 189$ | $4134 \pm 103$ |
| Between Jet Formation and Impact | | | | |
| $\langle \Delta x_j^{f\text{-}i} \rangle$ | (mm) | $240.1 \pm 5.3$ | $276.0 \pm 4.4$ | $284.1 \pm 2.8$ |
| $\langle \Delta y_j^{f\text{-}i} \rangle$ | (mm) | $41.6 \pm 1.6$ | $50.0 \pm 1.7$ | $53.4 \pm 1.5$ |
| $\langle \Delta t^{f\text{-}i} \rangle$ | (ms) | $148.9 \pm 2.6$ | $166.8 \pm 2.5$ | $169.7 \pm 2.2$ |
| $\langle a_{c,x}^{f\text{-}i} \rangle$ | (m/s$^2$) | -1.349 | -1.100 | -1.049 |
| $\langle a_{c,y}^{f\text{-}i} \rangle$ | (m/s$^2$) | -0.820 | -0.894 | -0.937 |
| $\langle a_{j,x}^{mj\text{-}i} \rangle$ | (m/s$^2$) | 1.030 | 1.059 | 1.281 |
| $\langle a_{j,y}^{mj\text{-}i} \rangle$ | (m/s$^2$) | -7.426 | -7.995 | -7.347 |
| Maximum Wave Crest Height | | | | |
| $\langle y_{\max} \rangle$ | (mm) | $107.8 \pm 0.3$ | $110.6 \pm 0.3$ | $111.5 \pm 0.5$ |

Table 2: Table of geometric and kinematic wave parameters for the three breakers at the times of jet formation and jet impact. The $\pm$ values with each average quantity indicate one standard deviation as measured from the profiles of each of the ten realizations of the breaker, while quantities without a $\pm$ value are measured directly from averaged breaker profiles. Most of the parameters are defined in figure 4. The time and $x$ and $y$ displacement of the jet tip from formation to impact are represented by $\Delta t^{f\text{-}i}$, $\Delta x_j^{f\text{-}i}$, $\Delta y_j^{f\text{-}i}$, respectively. The averaged $x$ and $y$ components of the accelerations of the wave crest and jet tip are represented by $a_{c,x}^{f\text{-}i}$, $a_{c,y}^{f\text{-}i}$, $a_{j,x}^{mj\text{-}i}$, $a_{j,y}^{mj\text{-}i}$ respectively. The maximum wave crest height relative to the still water level is given by $\langle y_{\max} \rangle$.

polynomials evaluated at $\tilde{t} = 0$ and the average speed $\langle V_j^i \rangle$ and angle $\langle \theta_{V_j^i} \rangle$ of this velocity for the three waves are given in Table 2. The jet tip impact speed increases monotonically by 5.6% (from 1.904 to 2.010 m/s) with breaker intensity increasing from the weak to strong breaker. The angle $\langle \theta_{V_j^i} \rangle$ increases from 21.9° to 27.6° as the breaker intensity increases from the weak to the strong breaker. It should be kept in mind that this is the angle made by the jet tip velocity vector; the angle of the front face of the jet as measured in a single

LIF image, $\theta_j^i$, at the time of jet impact is also included in the table. This angle is relatively independent of breaker intensity.

Brocchini & Peregrine (2001) used scaling arguments to predict the various types of disturbances of an air-water free surface produced by turbulent fluid motions. These surface disturbance types are identified in regions on a diagram with vertical axis, $q$, defined as a "turbulent velocity scale", and horizontal axis, $L$, defined as a length of the "most energetic turbulent scales", see their figure 10. There are no direct measurements of these quantities in the present experiments, but if $q$ is taken as the speed of the jet tip at impact, $\langle V_j^i \rangle$, and $L$ as the vertical height from the jet tip to the crest point at impact, $\langle r_y^i \rangle$, then the three points for the present breakers, see table 2, when plotted on figure 10 of Brocchini & Peregrine (2001) are all in the zone at the top of their plot, which is labeled "ballistic" and "splashing".

The vertical acceleration of the jet tip was determined by taking the second derivative of the above-described third order polynomials. It should be kept in mind that as second derivatives of measured trajectories, the accuracy of the accelerations is limited, approximately $\pm 1.0$ m/s². The average vertical component of the jet tip acceleration over the time span of the data, $\langle \overline{a}_{j,y} \rangle$, is found to be 7.43 m/s², 7.00 m/s² and 7.35 m/s² for the weak, moderate, and strong breakers, respectively. These accelerations are about 18% lower than the acceleration expected for a free falling object (see the curve from ballistic theory curve in figure 6(*a*)) and probably indicate the influence of surface tension and/or aerodynamic forces on the motion of the jet just before impact. Also, the jet tip in this study is a temporally evolving geometrical point on the curved jet tip surface, not the position of a particle of mass. The average horizontal component of the acceleration of the jet tip over the same period of time, $\langle \overline{a}_{j,x} \rangle$, is found to be 1.030 m/s², 1.059 m/s², 1.281 m/s² (in the direction of wave propagation, i.e., downstream) for the weak, moderate and strong breakers, respectively. The value of the horizontal acceleration in ballistic theory is, of course, equal to zero. Given the aforementioned accuracy figure, we cannot confidently identify any significant trend in the three vertical or horizontal acceleration values. The horizontal and vertical components of the distance traveled by the jet tip from formation to impact, $\Delta x_j$ and $\Delta y_j$, respectively, can be seen in figures 6(*b*) and are given in Table 2. Both distances increase monotonically with increasing breaking intensity.

As can be seen from Table 2, the variations in many of the above-described measured parameters are $\lesssim 10\%$ of their mean values over the three breakers. However, from qualitative observations of the LIF movies, one has the impression of a substantial increase in the scale and energy of the breaking region between the weak and the strong breaker, see Movie 2 and Movie 3 given in Supplementary Material. Also, as will be described in Part 2, there is a substantial increase in the number of droplets as the breaker strength is increased. Exceptions to the geometrical quantities that vary by small percentages are the vertical component of the velocity of the crest at the moment of jet formation (50% increase from the weak to the strong breaker), the vertical distance traveled by the jet tip (28% increase) and the parameters $r_x^i$, $r_y^i$ and $Q^i$, that describe the geometry of the crest region at the moment of jet impact and increase by approximately, 42%, 31% and 86%, respectively. The parameter $Q^i$ is the area under the upper surface of the plunging jet at impact, as shown by the colored area labeled $Q^i$ in figure 4. (It should be emphasized that $Q^i$ is a geometrical parameter that is defined using only the profile of the upper surface of the jet and wave crest. The relationship between $Q^i$ and the cross sectional area of the air tube entrapped under the jet at impact, $Q_{air}^i$, is likely to depend on the intensity of breaking as defined above and, through surface tension, the wavelength of the breaker. As noted in § 2.3, it is not possible to measure $Q_{air}^i$ with the present LIF technique. In lieu of measurements of $Q_{air}^i$, $Q^i$ will be used in Part 2 to correlate with the characteristics of the droplets generated by the breakers.) The increases in $r_x^i$ and $r_y^i$ are consistent with the increase in the upward vertical velocity of the crest, which is likely to indicate an increase in the vertical upward velocity of the jet tip as it is launched from the wave crest at $t = t^f$. This increase in vertical velocity might contribute to the increase in the horizontal distance traveled by the jet tip, resulting in impact with the wave face farther downstream where the water surface is lower. Plots of $\langle Q^i \rangle$ versus $\langle H^i/L^i \rangle$ at jet impact and $\langle v_c^f \rangle$ are given in figure 7(*a*) and (*b*), respectively. The plots indicate a nearly linear relationship (from only three data points) in both cases, however; the data in figure 7(*a*) conforms to the linear fit more closely.

### 3.3. *Wave Crest Profile Evolution after Jet Impact*

The behavior of the surface profile in the post jet-impact time period is presented and discussed in this subsection. Detailed measurements of profiles over the breaking crest are examined first in § 3.3.1 with profile sequences presented in a reference frame moving with the wave crest. This is followed by an examination of the profile measurements covering the entire measurement plane in § 3.3.2, presented in laboratory-fixed coordinates. In all cases, ensemble averages and standard deviations are computed from profile sequences that are spatially and temporally aligned, see § 3.1.

#### 3.3.1. *Breaking region development and evolution after jet impact*

After jet impact, significant run-to-run variations of the profiles are found in some spatial regions while in other regions the profiles are quite repeatable. The run-to-run variations are to be expected since at jet impact the flow begins transition to a temporally evolving and spatially nonuniform turbulent flow. An illustration of the run-to-run variation of the surface profiles after jet impact is given in figure 8(*a*) in which 10 profiles, one from each realization of the strong breaker, all recorded $\tilde{t} = 171 \pm 12.3$ ms after jet impact, are shown along with the ensemble average profile and a measure of standard deviation, $n_{sd}$, see figure 8(*b*) for definition. Starting from the right and proceeding to the left (downstream in laboratory coordinates), the profiles are at first well aligned; the first local height maximum is the original crest of the breaker and the second maximum is both the current wave crest point (the highest point on the profile) and the top of the first splash zone. The local minimum, referred to below as an indentation, between these two maximums is the point where the upper surface of the plunging jet is now submerging under the splash zone. Farther downstream of the indentation, the run-to-run variation in surface height increases with the maximum variation occurring near the leading edge (called the toe) of the splash zone (also called the turbulent breaking region). In the region downstream of the toe, the profiles are nearly the same in each run. It should be noted that the entire region shown in the plot is downstream of the jet impact location, $\tilde{x} = 0$.

Details of the evolution of the ensemble average profile and the standard deviation profile histories of the crest region for the weak, moderate and strong breakers are more clearly seen in the sequences of profiles in figure 9(*a*), (*b*) and (*c*), respectively. The plots are shown in a reference frame moving with the speed of the crest point at the moment of jet impact, $\langle u_c^i \rangle$ (see table 2), and cover a streamwise width of $0.4\lambda_0$ and a time interval of $0.55 f_0^{-1}$. Further details of the plots are given in the figure caption. In the moving reference frame of these plots, profile features that move faster (slower) than $\langle u_c^i \rangle$, move to the left (right) as time increases. The profiles in figures 9(*a*), (*b*) and (*c*) are colored locally according to the corresponding local value of $n_{sd}$. The horizontal and vertical positions of the indentations, the crest point, and the toe of the splash, all extracted from the profiles in figure 9, are plotted versus time ($\tilde{t} f_0$) in figure 10. Finally, for comparison, a profile history from a single realization of the strong breaker is given in figure 9(*d*). Comparison of the ensemble average (subplot(*c*)) and single realization (subplot(*d*)) profile histories helps to demonstrate the repeatability of the process, particularly the space-time locations of the indentations and splashes.

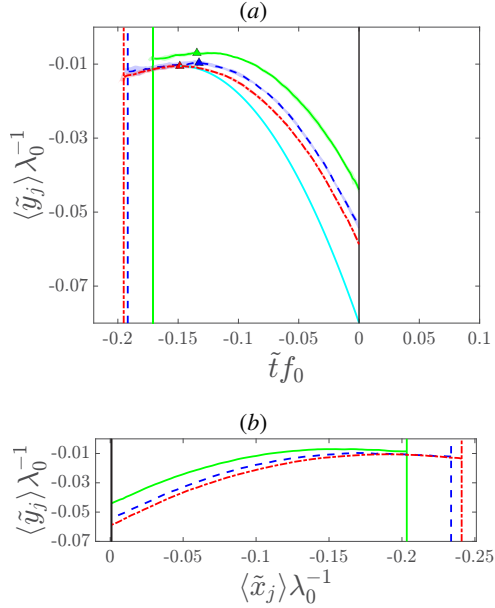

Figure 6: The ensemble average dimensionless jet tip height,$\langle \tilde{y}_j \rangle \lambda_0^{-1}$ is plotted versus $\tilde{t} f_0$, in subplot (*a*) and versus the ensemble average dimensionless jet tip horizontal position, $\langle \tilde{x}_j \rangle \lambda_0^{-1}$, in subplot (*b*). See the caption of figure 5 for the key to the colors and line types of the curves and vertical lines in the plots. In (a), the solid cyan line is the free-fall ballistic trajectory and the colored triangles indicate the location of the maximum jet tip height.

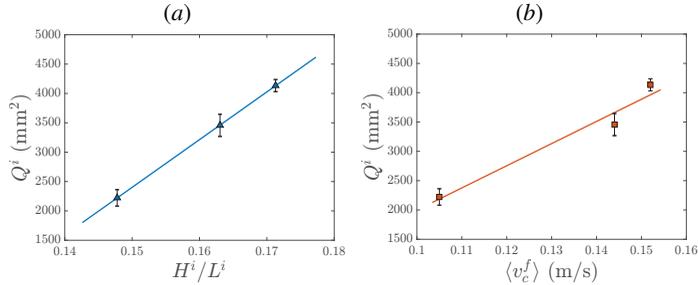

Figure 7: The estimated volume under the jet at the time of jet impact per unit length of crest, $Q^i$, is plotted vs. $\langle H^i/L^i \rangle$ and $\langle V_c^f \rangle$ in plots (*a*) and (*b*) , respectively. In (*a*) the solid straight line is a linear fit of the form $Q = p_1 * (H^i/L^i) + p_2$ where $p_1 = 81250$ and $p_2 = $ -9786, with $R^2 = 1$. In (*b*) the solid straight line is a linear fit of the form $Q = n_1 * \langle v_c^f \rangle + n_2$ where $n_1 = 37810$ and $n_2 = $ -1783, with $R^2 = 0.96$.

The repeatability of the early phase of the breaking process, including the jet impact and several cycles of splash generation and impact is demonstrated by the structure of each of the ensemble averaged profile histories as well as the qualitative similarity of the profile histories of the three breaking waves in figure 9. Despite the fact that profiles are taken from the period in which the flow transitions from a laminar to a turbulent flow, many features of the profile histories are repeatable enough in space and time in each realization to survive the ensemble averaging. The first repeatable feature is, of course, the average crest profile at the moment of jet impact, which is the first (lowest) profile. These are the same profiles of

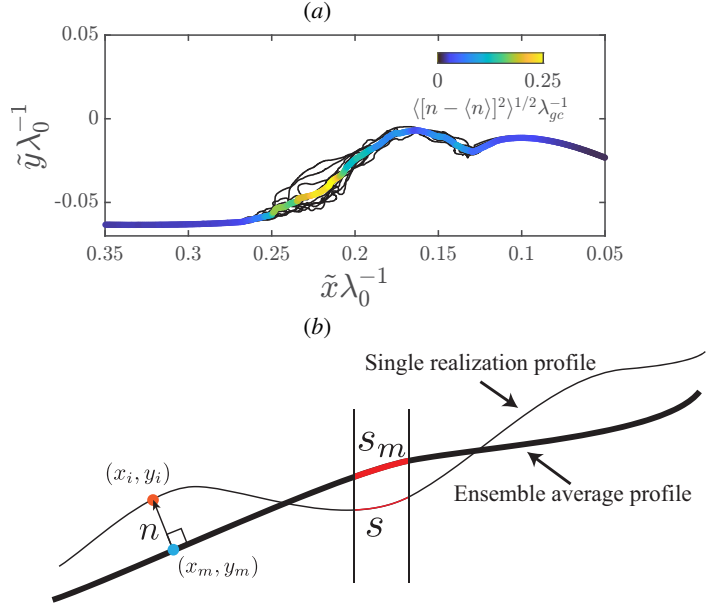

Figure 8: (*a*) Breaker profiles in $\tilde{y}\lambda_0^{-1} - \tilde{x}\lambda_0^{-1}$ coordinates from ten realizations of the strong breaker (solid black lines) all measured $171 \pm 12.3$ ms after jet impact. The thick colored line is the ensemble average profile, where the color contour along this line indicates the standard deviation of the normal distance $n$, $n_{sd} = \sqrt{\langle [n - \langle n \rangle]^2 \rangle}/\lambda_{gc}$, where $n$ is defined in (*b*) as the distance between the average profile and the individual profiles measured at each point along the average profile in the direction of its local normal and $\lambda_{gc} = 1.7$ cm, is the wavelength of the gravity-capillary wave with minimum phase speed. A second measure of profile variability, based on the profile local arc length, is also defined in (*b*) as the standard deviation of the difference between the local arc length of the ensemble average profile ($s_m$) and the corresponding local arc length (*s*) of the profile of a single realization of the breaker nondimensionalized by $s_m$, $s_{sd} = \sqrt{\langle [s/s_m - 1]^2 \rangle}$.

the crest region that are shown in figure 3(*d*). The $n_{sd}$ magnitude along the first profile is essentially zero for each of the three breakers. Immediately after impact, an indentation forms between the plunging jet's upper surface and the splash generated by the first impact. This indentation quickly slows down and connects to the zone of very high $n_{sd}$ (between profiles *iii* and *iv* and centered near horizontal position -0.2) that is associated with the bursting of bubbles entrapped under the plunging jet at the moment of impact, see Movie 4 included as supplemental material. The $n_{sd}$ level in this region increases monotonically with increasing breaking intensity, see § 3.3.2 for more details. The first indentation is also discussed in a number of studies including the experiments in Bonmarin (1989) and the 2D numerical computations in Iafrati (2009). In Bonmarin (1989), the idea of air entrainment via this indention is discussed along with a few surface profiles, while in Iafrati (2009), analysis of the surface profile contours show that the indentation is deep and closes near the surface, entrapping a pocket of air. The LIF method in the present study only records the full depth of the indentation if the laserlight rays are falling along the centerline of the cross-stream shape of the indentation and if the camera's line of sight is aligned with the indentation across the width of the tank. With a fixed camera, as in the present measurements, this alignment would only occur at one instant in time during a given breaking event. Since the cameras are stationary and not pointed across the tank width, the measured profiles indicate the location

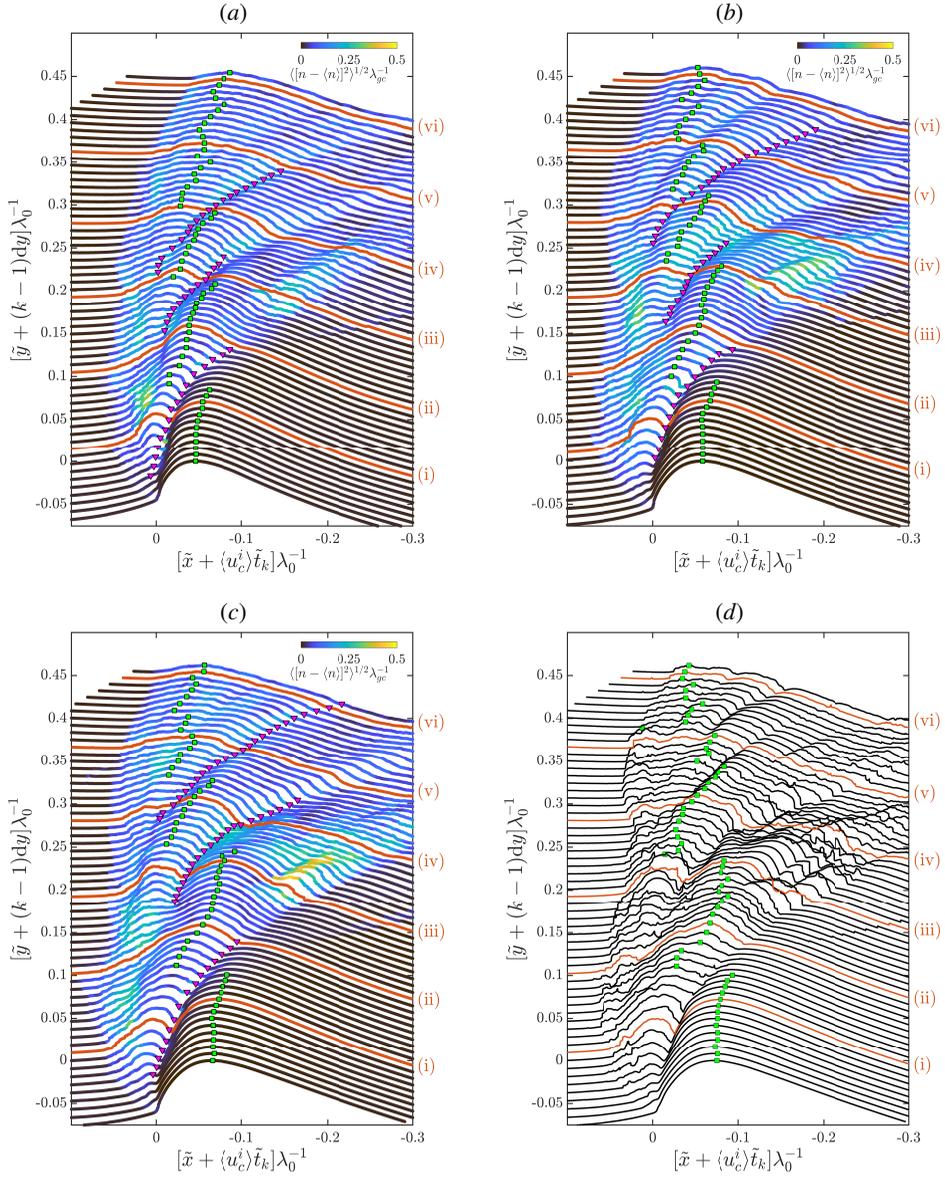

Figure 9: The evolution of the ensemble average crest profiles of the weak, moderate and strong breakers are presented in subplots (*a*), (*b*) and (*c*), respectively. Each line is the average profile at an instant in time, $\tilde{t}_k$, after jet impact ($k = 0$) and the color indicates the local value of the nondimensional standard deviation of the local normal distance, $n_{sd}$, see figure 8 its for definition. Each successive profile is plotted d$y = 10$ mm above the previous profile and the temporal separation between profiles is $\tilde{t}_{k+1} - \tilde{t}_k = 0.0123$ s. The breaker profile (i) occurs at $\tilde{t}_9 = 0.1107$ s and subsequent profiles marked by Roman numerals are 10 profiles apart

. The profiles are shown in a reference frame moving with the speed of the crest point at the moment of jet impact, $\langle u_c^i \rangle$. The green squares mark the location of the highest point on each profile and the magenta upside down triangles are the local minimums of height in the indentations. A similar set of surface height profiles, but for only one realization of the weakest breaker and plotted in the laboratory reference frame, was presented in Erinin *et al.* (2019). Subplot (*d*), is a profile evolution plot for one realization of the strong breaker.

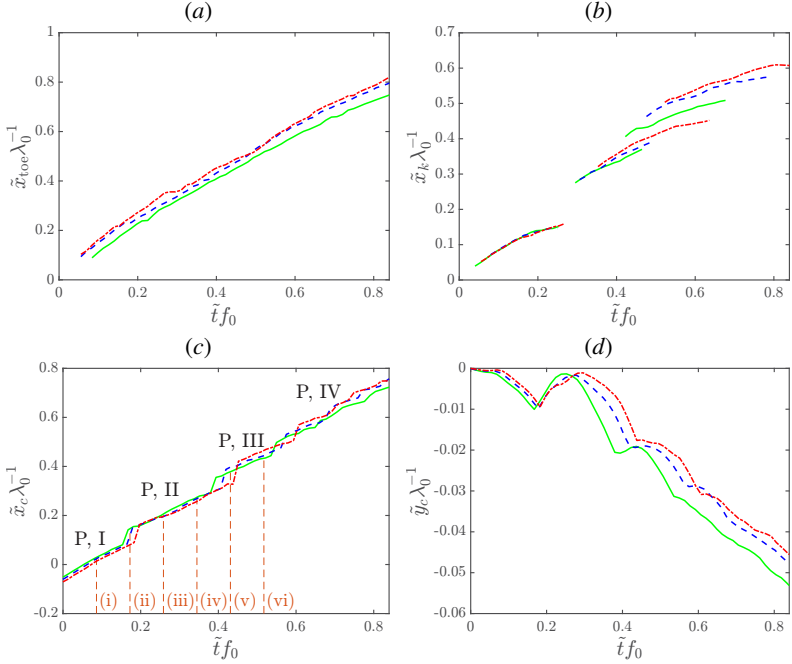

Figure 10: Plots of the dimensionless coordinates of several features of the breaking wave crest are plotted versus $\tilde{t}f_0$ in the laboratory reference frame. Subplots (*a*) and (*b*): The horizontal coordinates of the leading edge of the turbulent region (toe, $\bar{x}_{\text{toe}}\lambda_0^{-1}$) and the local minima of the three indentations ($\bar{x}_k\lambda_0^{-1}$), respectively. Subplots (*c*) and (*d*): The horizontal ($\bar{x}_c\lambda^{-1}$) and vertical ($\bar{y}_c\lambda^{-1}$) coordinates of the crest point, respectively. The time range in each plot is from the moment of jet impact to $\approx 730$ ms after jet impact, the same time interval as the profiles shown in figure 9. The vertical red dashed lines labeled (i) to (vi) in (*c*) are plotted at the times of the profiles in figure 9 with the same labels

of the indentation, but not the deep center of the cross section. In Part 2, it will be shown that a wall of droplets is ejected at the instant in time when the indentation disappears in the profile sequence and that this disappearance is the result of the rapid rise of the bottom of the thin nearly vertical air crater at the indentation site.

The splash generated by the plunging jet impact forms a region of high $n_{sd}$ but the motion of its leading edge, the toe, is highly repeatable, see the region of the splash zone between profiles (*i*) and (*ii*) and to the left of the first indentation in figure 9(*a*) to (*c*). Another two cycles of splash impact and generation are also visible with the second splash impact occurring between profiles (*ii*) and (*iii*) and the third between profiles (*iii*) and (*iv*). The impact-splash cycles are particularly prominent in the profiles from the two stronger breakers. After the initial jet impact, the leading edge of the turbulent breaking region accelerates upstream in all three profile histories and goes through cycles of accelerating and decelerating between profiles (ii) and (iv) as the splashes form and impact the wave face. The speed of the front splash region is close to the initial phase speed of the wave at several times, between surface profile (ii) and (v), but then decelerates by the time it reaches surface profile (vi). The horizontal position of the leading edge (toe) of the breaking region, $\langle\bar{x}_{\text{toe}}\rangle\lambda_0^{-1}$, is plotted versus $\tilde{t}f_0$ in figure 10(*a*). The toe position increases monotonically in time and, at any time, increases with breaker strength. There is evidence of some surging forward and backward in some of the curves, but the amplitude is small.

In a manner similar to the formation of the first indentation, with each splash impact an indentation forms between the plunging splash and the secondary splash that forms ahead of it. Each of these indentations at first moves at or near the speed $\langle u_c^i \rangle$ and quickly slows down and falls behind the crest. The streamwise positions of these indentations, $\langle \tilde{x}_k \rangle$, were extracted from the ensemble average profiles and are plotted in figure 10(*b*). The streamwise trajectory of the first indentation is nearly the same for the three breakers. The three trajectories of the second and third indentations are spread out in time and streamwise position, with the trajectories starting later in time and farther from the initial jet impact point as the breaker strength is increased.

The horizontal and vertical positions of the crest point (the highest point on each profile), $\langle \tilde{x}_c \rangle \lambda_0^{-1}$ and $\langle \tilde{y}_c \rangle \lambda_0^{-1}$, respectively, are plotted versus $\tilde{t} f_0$ in figures 10(*c*) and (*d*), respectively. The crest point is also depicted as the green squares in the profile history plots in figure 9. The three curves of horizontal position in subplot (*c*) follow a similar trend with regions of nearly linear increase connected by jumps downstream at $0.201 f_0^{-1}$, $0.437 f_0^{-1}$, and $0.575 f_0^{-1}$. Comparison with the profiles in figure 9(*a*), (*b*) and (*c*), indicates that the jumps in $\langle \tilde{x}_c \rangle \lambda_0^{-1}$ occur as the crest point jumps across an indentation from the main crest to the first splash, from the first to the second splash, etc. The jumps occur at increasingly later times as the breaker intensity is increased. If one imagines a smooth curve going through the data set for each wave, one can see that the curves for the three waves would be nearly the same, as are the raw data curves between each of the jumps. Also, the local curve slope is higher for the overall curve than for the regions between jumps. The speed of the crest point in each of the linear segments, which are labeled by uppercase Roman numerals in figure 10(*c*), are listed in table 3. These speeds are substantially less than $\langle u_c^i \rangle$ and $c_0 = 1.358$ m/s. The variable $\langle u^p/c \rangle$ in line one is the speed from a least squares fit of a straight line to the data for each breaker over the entire range of the plot. These speeds are only a little less than $c_0$. As seen in figure 10(*d*), the height of the crest point, $\langle \tilde{y}_c \rangle \lambda_0^{-1}$, decreases dramatically after jet impact, approximately $0.05 \lambda_0$ (a little more than 50% of the maximum crest height, see last line of table 2) in a time interval $\Delta \tilde{t} f_0^{-1} = 0.8$. It is thought that this decay is due partially to the conversion of wave energy to energy in the turbulence and due to the wave crest moving past the peak of the wave packet envelope. On top of this decaying curve is an oscillation with decaying amplitude; each local minimum occurs at the jump points in subplot (*c*), where the crest point moves from one ripple to the next. The overall slopes and shapes of the curves are similar in the three breakers, but the decay occurs later in time as the breaker intensity is increased. This general behavior of $x_c$ and $y_c$ is also seen in Bonmarin (1989) (figures 23, 24, 26 and 28) where the positions of the breaking crest and the first and second splash-ups are plotted versus time for a single plunging breaking event.

### 3.3.2. *Wave Field Evolution after Jet Impact*

In this subsection, the surface profile measurements over the entire measurement region in time ($-500$ ms $\leqslant \tilde{t} \leqslant 2,500$ ms) and space ($-0.3$ m $\leqslant \tilde{x} \leqslant 1.0$ m) are presented and discussed. Data for the ensemble averaged surface height, the local surface normal distance standard deviation ($n_{sd}$), and the local arc length standard deviation ($s_{sd}$) on the $\tilde{x} \lambda_0^{-1} - \tilde{t} f_0$ plane are shown in figures 11, 12 and 13, respectively. Each figure contains subplots for the weak, moderate, and strong breakers in (*a*), (*b*), and (*c*), respectively. The profile and $n_{sd}$ data in figure 9 are from a diagonal band (tilted toward increasing $\tilde{t} f_0$ and $\tilde{x} \lambda_0^{-1}$) starting at the jet impact location in the contour plots in figures 11 and 12. The definitions of the local surface normal distance ($n$) and the local arc length of the mean and instantaneous surface profiles ($s_m$ and $s$), respectively, are given in figure 8(*b*). These two measures of standard deviation were chosen to emphasize different features of the run-to-run variations of the profile shape, as is discussed below.

From the surface height contour maps shown in figure 11, the mean spatio-temporal evolution of the large-scale surface motion of the focusing wave packet can be analyzed. The height contour maps for the three breakers are quite similar. Perhaps the most important wave feature in these plots is the trajectory of the main breaking wave crest, the yellow band indicated by **A** and traveling from right to left as time increases (vertically up in the subplots). The sharp yellow-green transition on the lower side of the breaker crest is near the toe of the turbulent breaking region. In the moments leading up to jet impact, the mean wave crest height reaches a maximum, indicated by the green square. The jet tip impact point is denoted by the red triangle at $(\tilde{x}\lambda_0^{-1}, \tilde{t}f_0) = (0, 0)$, after which the surface becomes irregular as the wave breaks, see discussion in § 3.3.1. In laboratory coordinates, the jet tip impact point for the weak breaker is located at 5.93 m downstream from the back face of the wave maker, while the focusing distance parameter for the wave maker motion equation is $x_b = 6.17$ m, see equation 2.1. After the jet impact point, the breaking crest continues to decrease in amplitude while the amplitude of the crest following the breaker, indicated by **D** in the subplots, increases. This behavior is consistent with the expected evolution of a deepwater wavetrain moving at its phase velocity through an amplitude envelope that moves in the same direction at the group velocity. A third crest of smaller height is barely visible in the upper right corner of the subplots. Shortly after jet impact, three indentations form sequentially on the main wave crest. The indentations are seen as faint dark orange lines in the otherwise yellow region at the breaker crest and are identified by the three sets of three magenta downward triangles drawn beginning at the heads of the three arrows at **B** in the three contour plots. These features were discussed in reference to the crest profile plots in figure 9 where the indentations can be seen more clearly in the sequences of profiles shown in crest-fixed coordinates. The downward magenta triangles for each indentation are the first, midrange and last downward magenta triangles of the corresponding indentation in figures 9(a), (b) and (c).

| Variable | Unit | Weak | Moderate | Strong |
|---|---|---|---|---|
| $\langle u_c^P \rangle$ | (m/s) | 1.27 | 1.34 | 1.38 |
| $\langle u_c^I \rangle$ | (m/s) | 1.18 | 1.19 | 1.19 |
| $\langle u_c^{II} \rangle$ | (m/s) | 1.02 | 0.99 | 0.98 |
| $\langle u_c^{III} \rangle$ | (m/s) | 0.88 | 0.82 | 0.83 |
| $\langle u_c^{IV} \rangle$ | (m/s) | 1.08 | 1.21 | 1.16 |
| $\langle u_{toe} \rangle$ | (m/s) | 1.64 | 1.67 | 1.80 |

Table 3: Post impact horizontal speeds of the crest point and the toe of the three breakers. The speeds $\langle u_c^P \rangle$ are from the slopes of straight lines fitted to the entire range of data in figure 10(c). The speeds $\langle u_c^I \rangle$ to $\langle u_c^{IV} \rangle$ are from the slopes of straight lines fitted to the nearly linear segments of the $\langle \bar{x}_c \rangle - \tilde{t} f_0$ curves in figure 10(c). The horizontal speed of the splash front, $\langle u_{toe} \rangle$, is measured immediately following jet impact from $\tilde{t} = [0 \text{ to } 0.14]$ using the data shown in figure 10(e).

As mentioned above, two measures of the local standard deviation (SD) of the breaker profiles are presented herein: the normal distance standard deviation, $n_{sd} = \sqrt{\langle [n - \langle n \rangle]^2 \rangle}/\lambda_{gc}$, and the arc length standard deviation, $s_{sd} = \sqrt{\langle [s/s_m - 1]^2 \rangle}$. Both quantities are dimensionless and are defined in figure 8(b). The normal distance SD was used in the discussion of the crest profile evolution in § 3.3.1. The normal distance was chosen rather than the vertical height as a measure of the difference between two water surface profiles because the height change exaggerates the miss match of the two curves when a small phase difference occurs between them at sharp profile features like the bottom of the indentations discussed in § 3.3.1.

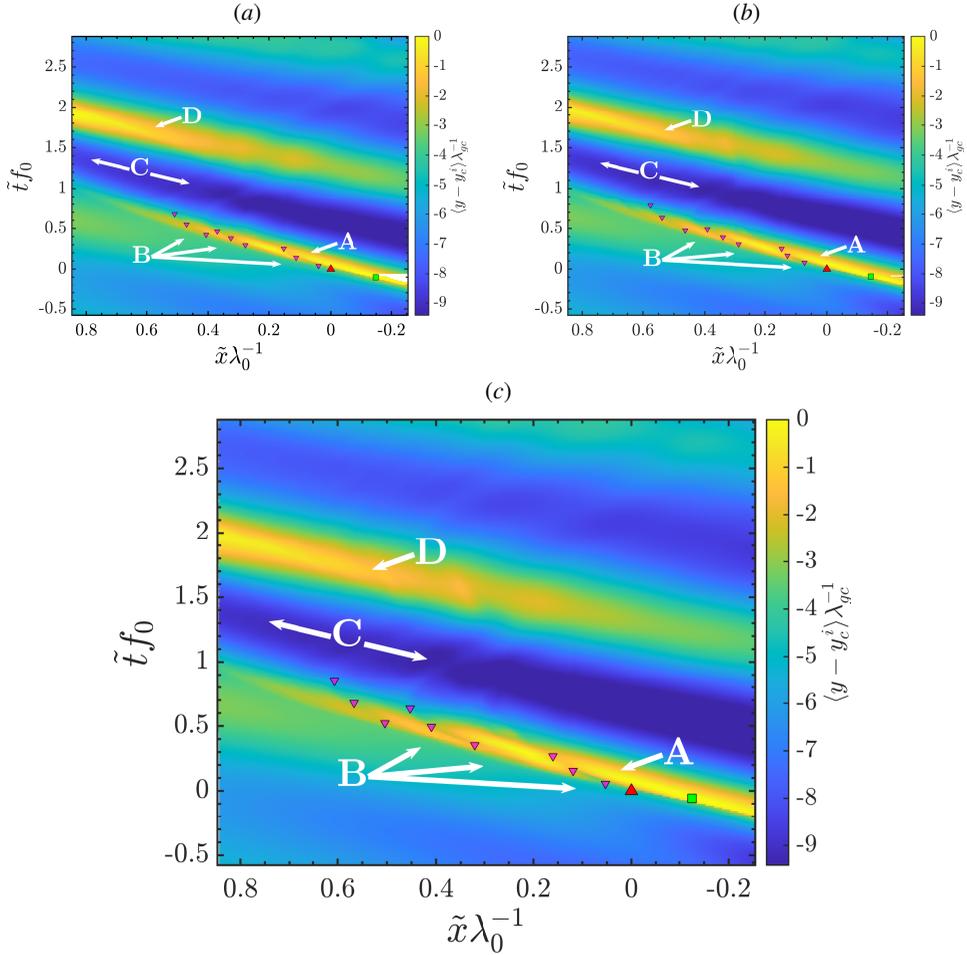

Figure 11: Contour maps of the nondimensional mean surface height $\langle \bar{y} \rangle \lambda_{gc}^{-1}$ from 10 breaker realizations on the $\tilde{x}\lambda_0^{-1} - \tilde{t}f_0$ plane for the weak, moderate and strong breakers are given in subplots $(a)$, $(b)$, and $(c)$, respectively. The wavelength $\lambda_{cg} = 1.71$ cm is the wavelength of the gravity-capillary wave with minimum phase speed. The contour maps are shown in the laboratory reference frame for the full measurement region, $\approx 1300$ mm in streamwise distance and $\approx 3000$ ms in time and have a spatial and temporal resolution of $\Delta x = 0.165$ mm and $\Delta t = 12.3$ ms, respectively. The position and time of the maximum height of the wave crest point and the jet tip impact are indicated by the green square and the red triangle, respectively. The call-outs **A-E** and the magenta downward triangles are discussed in the text. During the time approaching the maximum crest elevation, the back face of the breaking wave crest becomes partially obscured from the camera view and is not measured, as indicated by the white region just to the right of the green square in subplots (a) and (b).

The arc length SD is a measure of the relative shape of two profiles. If the individual and average profiles have the same shape and orientation over $\Delta x$, the difference in $s_{sd}$ is zero no matter the displacement $(\overline{n})$ between the two curves. For the same curves, $n_{sd}$ would equal the displacement. If the shapes of the two curves are different over $\Delta x$ but the curves are on top of one another, the $n_{sd}$ may be small, while $s_{sd}$ can be large. In order to explore this difference numerically, a plot of these two quantities for an interval with an average profile consisting of a straight line (say $y = 0$) and an individual profile consisting of a sine wave, $y = a \sin kx$, where $k = 2\pi/\lambda$ and $\lambda$ is the wavelength, is given in figure 14. The arc length and $\overline{n}$ are computed over the interval $0 \leqslant x \leqslant \lambda/2$. In this case, $\overline{n}$ is due to the shape of the profile within the interval $\Delta x$ and $\overline{n} = 2a/\pi$. As can be seen from the figure, $\overline{n}$ varies linearly with $ak$ while the arc length at first increases more slowly than $\overline{n}$ but soon increases more rapidly. Thus, compared to $n_{sd}$, $s_{sd}$ is less sensitive to small slope disturbances and more sensitive to large slope disturbances. Thus, these differences in the behavior of the two measures of standard deviation create contour plots that emphasize different features of the surface roughness.

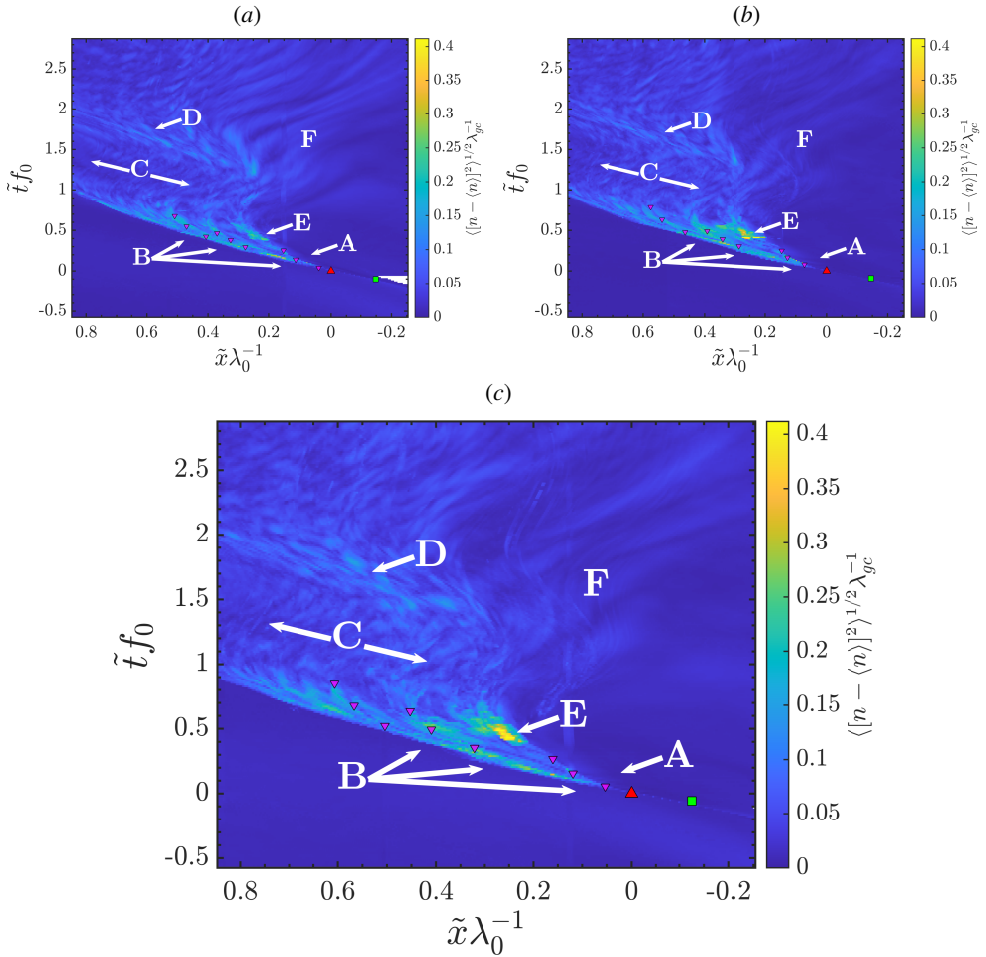

Figure 12: Contour maps of the nondimensional standard deviation of the local surface normal distance $n_{sd}$ generated from 10 realizations on the $\tilde{x}\lambda_0^{-1} - \tilde{t}f_0$ plane for the weak, subplot (*a*), moderate, subplot (*b*), and strong, subplot (*c*), plunging breakers. See figure 8(*b*) for definitions of $n_{sd}$. Plot details are given in the caption to figure 11.

The contour plots of $n_{sd}$ and $s_{sd}$ are given in figures 12 and 13, respectively. In these plots, the call-outs *A* through *D*, the points of maximum crest height and jet impact, and the downward magenta triangles marking the locations of the indentations are identical to those in the corresponding subplots in figure 11. The most striking feature of all of the $n_{sd}$ and $s_{sd}$ contour plots is the sharp boundary between the low-SD-level region in the lower left (downstream, i.e., in the direction of wave travel, of the breaking crest, ) and the high-SD-level region of the front face of the wave crest. This boundary is slightly ahead, down and to the left, of the sharp blue-yellow transition in the subplots in figure 11 and corresponds to the leading edge (toe) of the breaking region. The same boundary is visible in the $n_{sd}$ color contours on each of the the ensemble average profiles shown in crest-fixed coordinates in figure 9. The lower left regions of the $n_{sd}$ and $s_{sd}$ contour plots correspond to the water surface upstream of the breaker and the low values of $n_{sd}$ and $s_{sd}$ in these regions are a testament to the repeatability of the ten realizations of the non-breaking sections of the wave profile in $\tilde{x} - \tilde{t}$ coordinates.

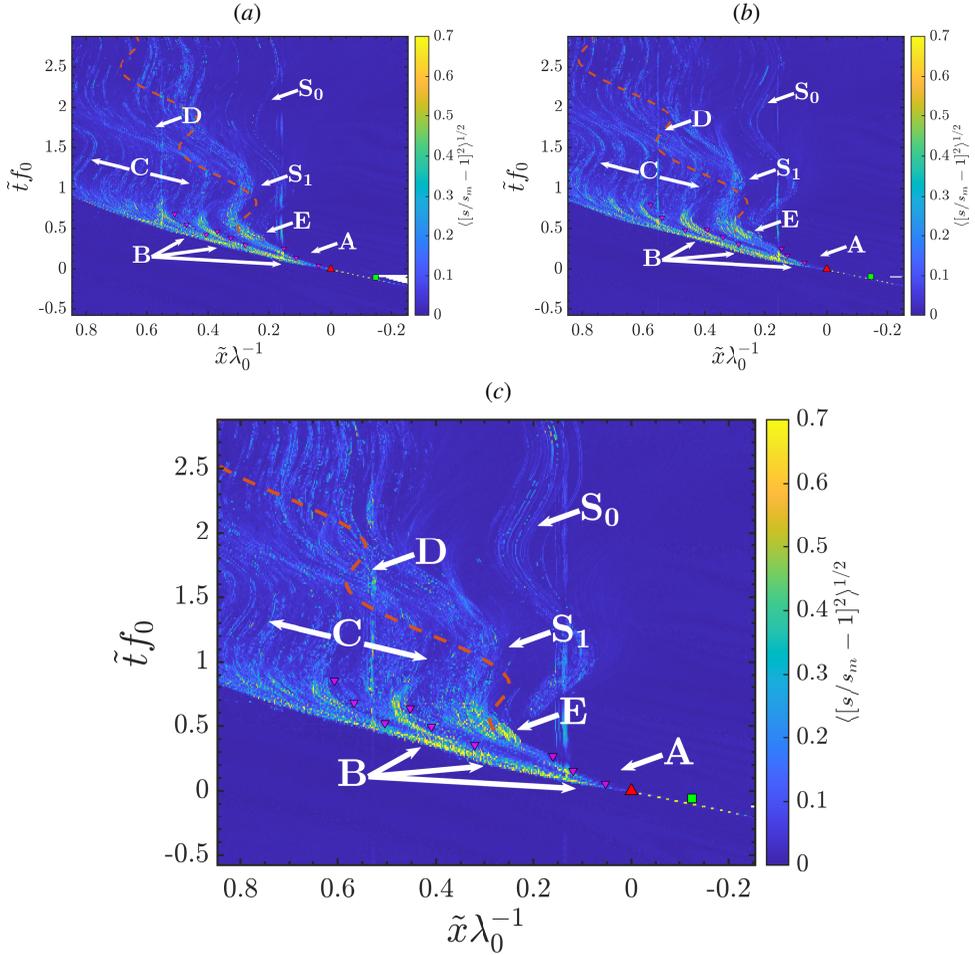

Figure 13: Contour maps of the nondimensional standard deviation of the local arc length $s_{sd}$ generated from 10 realizations on the $\tilde{x}\lambda_0^{-1} - \tilde{t}f_0$ plane for the weak, subplot ($a$), moderate, subplot ($b$), and strong, subplot ($c$), plunging breakers. See figure 8($b$) for definitions of $s_{sd}$. The dashed orange line in all three plots is the horizontal trajectory of a surface particle according to linear wave theory modified to include a large drift current, see text for discussion. The spatial resolution is $\Delta x = 2.740$ mm, other plot details are given in the caption to figure 11.

The $n_{sd}$ contour plots in figure 12 show a number of interesting features. In the case of the weak breaker, subplot ($a$), one can see three diffuse bright elongated regions along the locations of the three indentations in the height contour plots at positions $B$. The bright regions along the indentation locations become less distinct as the breaker intensity increases, see subplots ($b$) and ($c$). A small region of high $n_{sd}$ contour levels is found at location $E$, in all three subplots. This location is in line with but well past the end of the first indentation. The $n_{sd}$ levels in these small regions, approximately 2.07, 3.20, 3.43 mm in the weak, moderate and strong breakers, respectively, are the highest in each contour map. These dimensional values correspond to $0.025H^i$, $0.036H^i$ and $0.038H^i$ and $0.039\langle r^i_y \rangle$, $0.049\langle r^i_y \rangle$ and $0.049\langle r^i_y \rangle$, for the weak, moderate and strong breakers, respectively. See figure 4 for the definitions and table 2 for values of $H^i$ and $\langle r^i_y \rangle$ for each breaker. Observations of the diffuse-light movies indicates that the patch of high SD level corresponds to the location where the large air bubbles entrapped under the jet at impact escape from the back face of the wave well past the end of the trajectory of the first indentation, see Movie 4 given as supplemental material. The closing of the indentation crater, as discussed above, occurs well before the emergence of the bubbles entrapped under the plunging jet. This large bubble bursting region was also noted by other investigators in experiments, including Bonmarin (1989) and Blenkinsopp & Chaplin (2007), and in numerical calculations, including Wang *et al.* (2016) and Mostert *et al.* (2022).

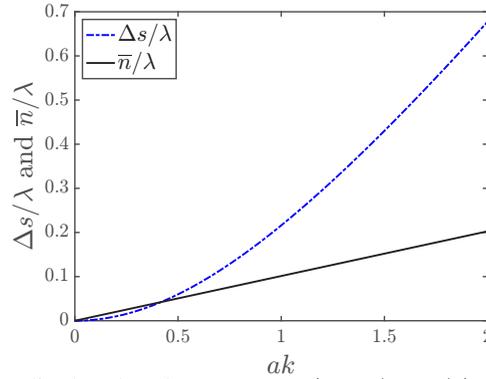

Figure 14: The normalized arc length parameter $\Delta s/\Delta x = (s - s_m)/\Delta x$ and the normalized average normal displacement $\bar{n}/\Delta x$ both over a measurement interval $\Delta x$, see figure 8 for definitions, with the mean profile taken as a straight horizontal line ($y = 0$, length $s_m = \Delta x$) and an individual profile realization taken as a sine wave, $y = a \sin kx$, where $k = 2\pi/\lambda$ and $\lambda$ is the wavelength. The measurement interval $\Delta x$ is from $x = 0$ to $x = \lambda/2$. The slope of the $\bar{n}/\Delta x$ curve is $1/\pi^2 = 0.101$, while the slope of the $\Delta s/\Delta x$ curve is zero at $ak = 0$ and asymptotes to a strait line with a slope of approximately 0.52 at $ak = 2.0$.

In the remaining areas of the $n_{sd}$ contour maps, the $n_{sd}$ level is lower than in the breaking crest region but exhibits some interesting well-organized features. The magnitude of $n_{sd}$ decreases in the trough ($C$) between the breaking crest and following wave crest, increases again at the following wave crest ($D$) to a lower level than on the breaker crest and falls off again after the second crest. The pattern of light blue small-scale features in the area bounded by the breaking crest on the lower left edge and approximately by the red dashed wavy line between $E$ and $D$ in the $s_{sd}$ contour plots in figure 13 is likely the manifestation of the breaker-generated turbulent flow. The slow downstream mean motion of the wavy boundary is consistent with the idea of turbulent vortical flow features moving with the fluid rather than the wave crest. The relatively high level of $n_{sd}$ at the following wave crest may be associated with an interaction between the flow field of the long wavelength carrier wave of the packet and the short waves generated by the turbulence, (Miller *et al.* 1991; Phillips 1981; Longuet-Higgins 1987), or the modification of the turbulence by the carrier wave flow field (Tucker & Reynolds 1968; Reynolds & Tucker 1975; Kevlahan & Hunt 1996, 1997; Hadzic *et al.* 2001). A larger scale faint wavelike structure is visible in the upper right portion of the plot ($F$) where as many as five wavelengths measuring approximately 120 mm in wavelength can be seen. This wavy feature is clearly visible in the plot for the weak breaker. The source of these waves is unknown, but may be associated with the interaction of the waves with the tank walls.

Contour plots of $s_{sd}$, are shown in figure 13. The bright regions of the $s_{sd}$ contour plots are located in the same general areas as those in the $n_{sd}$ contour plots and consist of many small bright spots and short streaks. Some of the highest average contour levels are at the leading edge of the breaking region, just to the left of the each of the three indentations and at the location of large bubble bursting that was also prominent in the $n_{sd}$ contour maps. The faint vertical lines at $\tilde{x}\lambda_0^{-1} \approx 0.18$ and 0.55 are due to the slight miss-matching of the profiles at the boundaries of the three images from which each profile is extracted. The local arc length seems to be particularly sensitive to errors in profile continuity at these boundaries. The collections of bright spots and streaks form striations with swaying motion. The three brightest of these striations issue from the three indentations (the first (right most) is labelled $S1$), and striations with similar shape and lower intensity are constantly produced all along the breaking crest region. The striations persist for the entire time span of the plot, about 2.7 wave periods, and sway to the left on the back face of the two wave crests and to the right on the front face of the following wave crest. With the exception of the striation marked by the call-out $S_0$ and issuing from approximately the end of the rightmost arrow at call-out $B$, there is, on top of the swaying motions, a general motion downstream (to the left). A visual inspection of some of the LIF images indicates that many of the bright spots and streaks that make up these striations are signatures of short wavelength ripples and individual bubbles that are in the light sheet during one or more frames of one or more of the LIF movies. These signatures seem to be strong enough to survive the ensemble averaging. This supports the hypothesis that the striations are collections of signatures of surface features generated at the breaking crest and moving downstream with the water surface particle velocity. The faint rightmost striation ($S_0$), which has only a small downstream drift, is thought to be in the region upstream of the breaking zone where the surface drift is caused primarily by nonlinear (nonbreaking) wave motion. On the other hand, the surface drift affecting the surface features making up the other striations is thought to include a strong component due to the intense surface current created by the breaking process. The surface features downstream (toward the lower left in the plot) of the breaking zone may be generated by the impacts of droplets ejected with downpstream trajectories from the breaking zone, as seen, for example, in Movie 4 given in Supplementary Material.

To explore the wavy striations in the $s_{sd}$ contour plots further, the trajectory given by

$$\tilde{x} = \tilde{x}_0 + a \cos\left(\omega(\tilde{t} - \tilde{t}_0)\right) + U_d(\tilde{t} - \tilde{t}_0) \tag{3.1}$$

was fitted by eye to the first striation ($S_1$) affected by a strong drift current in the three $s_{sd}$ contour plots. In this equation, the first term is a spatial phase shift, the second term mimics the orbital motion of the waves (with the carrier wave frequency $\omega$, amplitude $a$ and time shift $t_0$), and the third term mimics the motion due to a surface drift of constant speed $U_d$. The values of $x_0$, $t_0$, $a$, $T = 2\pi/\omega$ and $U_d$ from the qualitative fitting are given in table 2 and its caption and the resulting trajectories are plotted as the dashed wavy red curves in the three subplots of figure 13. The trajectories are qualitatively a good fit to the striation labeled $S1$ in each subplot, and the striations further downstream (fitted curves not shown), for at least the first wave period after breaking. In later time, the striations seem to drift downstream at a rate slower than $U_d$. The computed values of $U_d$, 0.13, 0.18 and 0.21 m/s for the weak, moderate and strong breakers, respectively, are substantial and approximately consistent with the fluid speeds in the wakes of breaking waves as reported in Duncan (1981, 1983), Rapp & Melville (1990) and Melville *et al.* (2002). The coincidence of these trajectories and the patterns in the contour plots lend further support to the idea that the arc length standard deviation is responding to small surface features, like bubbles, that move along with the surface drift current. According to the fitting, this drift current increases with breaker strength.

|  | Weak | Moderate | Strong |
|---|---|---|---|
| $t_0$ (s) | 0.52 | 0.52 | 0.54 |
| $x_0$ (m) | 0.35 | 0.35 | 0.33 |
| $U_d$ (m/s) | 0.13 | 0.18 | 0.21 |

Table 4: Parameters obtained by qualitative (visual) fitting of the the the drift velocity equation (3.1) to the arc length standard deviation contour plots in figure 13. The wave period, $T = 1.0$ s, and wave amplitude, $a = 0.07$ m, used in the fitting function were the same for all three waves. The results are plotted as the wavy dashed red line in each figure.

Comparisons of the magnitudes of $n_{sd}$ and $s_{sd}$ from one wave to another are also of interest. In the case of $n_{sd}$, the comparison is difficult because the contour maps consists of large fields of low-level signal with a few isolated areas of high-level signal. Average values over the entire measurement field or even over only the breaking wave crest have yielded very low levels that are not consistent with the qualitative appearance of the contour maps. Here, only a simple quantitative measure of $n_{sd}$ is chosen, a stable peak value obtained by averaging over a small region centered at the location with the maximum value, marked by $E$ in each subplot. This area, measuring $0.156 f_0^{-1}$ (135.3 ms) by $0.042 \lambda_0$ (49.7 mm), essentially covers the yellow contour level region in the plot for the strong breaker, subplot ($c$). The local maximum values obtained in this way are called $(n_{sd})_{max}$. Since the high $s_{sd}$ regions in the subplots in figure 13 cover a large area, an average value is calculated over the large rectangle defined by the corners $(\tilde{x}\lambda^{-1}, \ \tilde{t}f_0) = (0, 0)$ and $(0.85, 1.0)$. This average value is given the symbol $(s_{sd})_{ave}$. Plots of $(n_{sd})_{max}$ and $(s_{sd})_{ave}$ versus $Q^i$ are given in figure 15($a$) and ($b$), respectively, where the three data points in each plot appear to be form straight lines with positive slope, see figure caption for details. It should be noted that $(n_{sd})_{max}/\langle r_y^i \rangle$ and $(s_{sd})_{ave}$ vary from 0.034 to 0.056 and from 0.0538 to 0.064, respectively, from the weak to the strong breaker.

## 4. Conclusions

An experimental study of the temporal evolution of the profiles of plunging breaking waves created with mechanically generated dispersively focused wave packets was presented. The profiles were measured with a cinematic LIF method with high temporal and spatial resolution (650 pps and 180 μm/pixel, respectively) over a time period of 3.4 average wave periods and a streamwise position interval of 1.1 nominal breaker wave lengths. Three waves were studied and these waves were generated with wave maker motions with nearly the same shape (average frequency, 1.15 Hz) but different overall amplitudes. The three amplitudes were chosen to create breakers ranging from a weak plunging breaker, such that any reduction in amplitude created a spilling breaker, and a strong plunging breaker, such that any increase in amplitude created a breaker at a position at least one wavelength closer to the wave maker than the intended break point. These wave maker amplitudes differ by only approximately 8% of their average. Each of the three breakers was generated and measured ten times.

The water surface profiles extracted from the LIF images were first used to analyze the crest profiles during the period from the moment when the plunging jet starts to form to the moment when the plunging jet tip hits the front face of the wave. The profiles during this time period are exceedingly repeatable, save for a $\pm 1.5$ cm run-to-run variation in the streamwise position of jet tip at impact. It was found that while many of the measured parameters, including the crest height, the horizontal speed of the highest point on the wave profile (called herein the crest point), and the jet tip velocity just before impact differ by less than 10% between the weakest and strongest breakers, several changed more substantially. The variables with more dramatic ranges include the vertical velocity of the crest point at the moment of jet formation, 45%, the horizontal and vertical distances between the jet impact point and the crest point, approximately 35%, and a measure of the area under the plunging jet at the moment of impact, 86%.

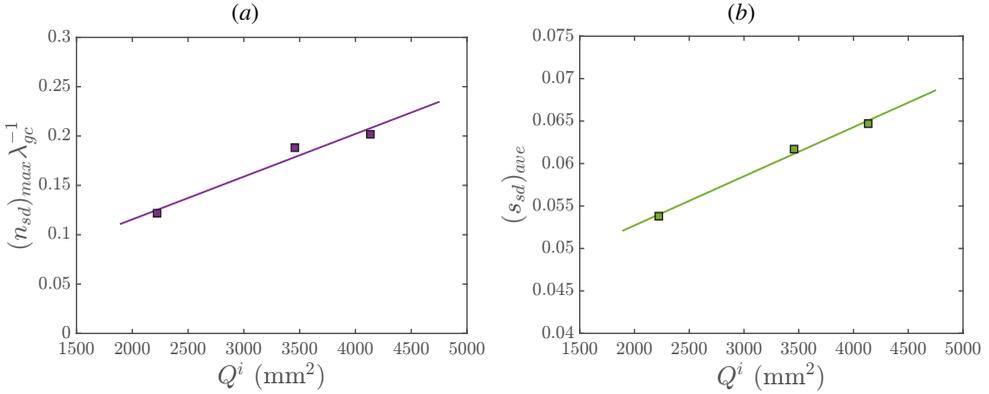

Figure 15: Plots of $(n_{sd})_{max}$ and $(s_{sd})_{ave}$ versus $Q^i$ are given in (*a*) and (*b*), respectively. The maximum value $(n_{sd})_{max}$ is the average value $n_{sd}$ in a region that is $11\Delta t = 135.3$ ms by $301\delta x = 49.7$ mm centered at the peak value of $n_{sd}$ in the large bubble popping region on the back face of the breaker. The average value $(s_{sd})_{ave}$ is the average of $s_{sd}$ over a rectangular region defined by the corners $(\bar{x}\lambda^{-1}, \tilde{t}f_0) = (0, 0)$ and $(0.85, 1.0)$. The solid line in each plot is a least-squares fit of a straight line to the three data points: $(n_{sd})_{max} = 4.33 \times 10^{-5}Q^i + 0.0289$ in (*a*) and $(s_{sd})_{ave} = 5.79 \times 10^{-6}Q^i + 0.0411$ in (*b*).

Immediately after jet impact, the flow begins the transition to turbulent flow. To explore this phase of breaking, the 10 profile sequences of each wave were first aligned relative to the time and position of jet impact and then, at each instant in time, the ensemble average profile and two measures of the distribution of standard deviation along the average profiles were computed. The first measure of standard deviation is the normal distance between each segment of the mean profile and the corresponding segments of each individual profile and the second is the difference between the arc length of corresponding segments of the mean and individual profiles. It is found that during this transition process there is a highly repeatable motion that appears in the ensemble average profile series. The motion consists of the first jet impact and subsequent splash up, followed by two cycles of splash impact and secondary splash generation. At each impact site a prominent indentation appears. These structures at first move with the speed of the crest but quickly slow down as the wave crest moves through the self-generated turbulent flow field. Not surprisingly, this behavior is similar to that reported for gentle spilling breakers, Duncan *et al.* (1999*b*). From the LIF and diffuse light movies of the breaking events, it can be seen that the large air bubbles entrapped under the plunging jet at impact come to the surface and burst at a position and time just past the end of the first indentation, which at this point is on the back face of the breaking wave crest.

The $\bar{x}$-$\tilde{t}$ contour maps of the normal distance standard deviation have a very low overall average while local regions of high standard deviation are found at the leading edge of the turbulent breaking region, just upstream of the indentations, and the above-described location of the bursting of the large bubbles entrained under the plunging jet. The local average of peak value of $n_{sd}$ at the latter location increases from 3.4 to 5.6 percent of the vertical distance from the jet impact location to the wave crest from the weak to the strong breaker. The arc length is sensitive to sharply curved regions of the surface created by small waves and bubbles. The resulting $x$-$t$ contour map reveals a wavy pattern induced by oscillatory fluid particle motion in the focused wavetrain accompanied by a strong downstream drift probably created by the drift motion of the highly nonlinear waves and by the surface current induced by breaking. The contour maps of the arc length standard deviation are more evenly distributed and the average value over the breaker was computed. It is found that this average increases from 5.4 to 6.5 percent from the weak to the strong breaker.

**Acknowledgements.** The authors thank undergraduate students Benjamin Schaefer, Benjamin Cha and Nicholas Lawson for help with the image processing related to these experiments.

**Funding.** The support of the Division of Ocean Sciences of the National Science Foundation under Grants OCE0751853, OCE1829943 and OCE1925060 is greatly acknowledged.

**Declaration of interests.** The authors report no conflict of interest.

REFERENCES

BANNER, M L & PEREGRINE, D H 1993 Wave breaking in deep-water. *Ann. Rev. Fluid Mech.* **25**, 373–397.

BLENKINSOPP, C.E & CHAPLIN, J.R 2007 Void fraction measurements in breaking waves. *Proceedings of the Royal Society A: Mathematical, Physical and Engineering Sciences* **463** (2088), 3151–3170.

BONMARIN, P 1989 Geometric-properties of deep-water breaking waves. *J. Fluid Mech* **209**, 405–433.

BROCCHINI, M. & PEREGRINE, D. H. 2001 The dynamics of strong turbulence at free surfaces. part 1. description. *J. Fluid Mech.* **449**, 225–254.

CHEN, G, KHARIF, C, ZALESKI, S & LI, J 1999 Two-dimensional navier-stokes simulation of breaking waves. *Phys Fluids* **11** (1), 121–133.

COKELET, E. D. 1977 Steep gravity-waves in water of arbitrary uniform depth. *Philosophical Transactions of the Royal Society A-Mathematical Physical and Engineering Sciences* **286** (1335), 183–230.

DEIKE, L., MELVILLE, W. K. & POPINET, S. 2016 Air entrainment and bubble statistics in breaking waves. *J. Fluid Mech.* **801**, 91–129.

DEIKE, LUC, PIZZO, NICK & MELVILLE, W. KENDALL 2017 Lagrangian transport by breaking surface waves. *J. Fluid Mech.* **829**, 364–391.

DEIKE, LUC, POPINET, STEPHANE & MELVILLE, W. KENDALL 2015 Capillary effects on wave breaking. *J. Fluid Mech.* **769**, 541–569.

DERAKHTI, MORTEZA & KIRBY, JAMES T 2014 Bubble entrainment and liquid–bubble interaction under unsteady breaking waves. *J. Fluid Mech.* **761**, 464–506.

DERAKHTI, MORTEZA & KIRBY, JAMES T. 2016 Breaking-onset, energy and momentum flux in unsteady focused wave packets. *J. Fluid Mech.* **790**, 553–581.

DRAZEN, DAVID A. & MELVILLE, W. KENDALL 2009 Turbulence and mixing in unsteady breaking surface waves. *J. Fluid Mech.* **628**, 85–119.

DRAZEN, DAVID A, MELVILLE, W KENDALL & LENAIN, LUC 2008 Inertial scaling of dissipation in unsteady breaking waves. *J. Fluid Mech.* **611**, 307–332.

DUNCAN, J. H. 1981 An experimental investigation of breaking waves produced by a towed hydrofoil. *Proc. Roy. Soc. of Lond., A,* **377**, 331–348.

DUNCAN, J. H. 1983 The Breaking and Nonbreaking Wave Resistance of a Two-Dimensional Hydrofoil. *J. Fluid Mech.* **126**, 507–520.

DUNCAN, J. H., QIAO, H., PHILOMIN, V. & WENZ, A. 1999*a* Gentle spilling breakers: crest profile evolution. *J. Fluid Mech.* **379**, 191–222.

DUNCAN, J. H., QIAO, H. B., PHILOMIN, V. & WENZ, A. 1999*b* Gentle spilling breakers: crest profile evolution. *J. Fluid Mech.* **379**, 191–222.

ERININ, MARTIN A., WANG, SOPHIE D., LIU, REN, TOWLE, DAVID, LIU, XINAN & DUNCAN, JAMES H. 2019 Spray generation by a plunging breaker. *Geophysical Research Letters* **46**, 8244–8251.

HADZIC, I., HANJALIC, K. & LAURENCE, D. 2001 Modeling the response of turbulence subjected to cyclic irrotational strain. *Physics of Fluids* **13** (6), 1739–1747.

IAFRATI, A. 2009 Numerical study of the effects of the breaking intensity on wave breaking flows. *J. Fluid Mech.* **622**, 371–411.

KEVLAHAN, NKR & HUNT, JCR 1997 Nonlinear interactions in turbulence with strong irrotational straining. *J. Fluid Mech.* **337**, 333–364.

KEVLAHAN, N. K. R. & HUNT, J. C. R. 1996 Nonlinear interactions in turbulence with strong irrotational straining. In *Advances in Turbulence VI* (ed. S Gavrilakis, L Machiels & PA Monkewitz), *Fluid Mechanics and Its Applications*, vol. 36, pp. 239–242. EUROMECH; ERCOFTAC; COST, 6th European Turbulence Conference, Swiss Fed Inst Technol, Lausanne, Switzerland, JUL 02-05, 1996.

KIGER, K. T. & DUNCAN, J. H. 2012 Air-entrainment mechanisms in plunging jets and breaking waves. *Ann. Rev. Fluid Mech.* **44**, 563–596.

LAMARRE, E. & MELVILLE, W. K. 1991 Air entrainment and dissipation in breaking waves. *Nature* **351** (6326), 469.

DE LEEUW, G., ANDREAS, E. L., ANGUELOVA, M. D., FAIRALL, C. W., LEWIS, E. R., O'DOWD, C., SCHULZ, M. & SCHWARTZ, S. E. 2011 Production flux of sea spray aerosol. *Reviews of Geophysics* **49** (2).

LIN, C. & HWUNG, H. H. 1992 External and internal flow-fields of plunging breakers. *Experiments in Fluids* **12** (4-5), 229–237.

LONGUET-HIGGINS, M. S. 1976 Breaking waves in deep or shallow water. *Proceedings of the 10th Symposium on Naval Hydrodynamics* pp. 597–605.

LONGUET HIGGINS, M. S. 1984 New integral relations for gravity-waves of finite-amplitude. *J. Fluid Mech.* **149** (DEC), 205–215.

LONGUET-HIGGINS, M. S. 1987 The propagation of short surface-waves on longer gravity-waves. *J. Fluid Mech.* **177**, 293–306.

LONGUET-HIGGINS, M. S. 1995 On the disintegration of the jet in a plunging breaker. *J. Phys. Ocean.* **25** (10), 2458–2462.

LUBIN, P. & GLOCKNER, S. 2015 Numerical simulations of three-dimensional plunging breaking waves: generation and evolution of aerated vortex filaments. *J. Fluid Mech.* **767**, 364–393.

LUBIN, PIERRE, VINCENT, STEPHANE, ABADIE, STEPHANE & CALTAGIRONE, JEAN-PAUL 2006 Three-dimensional large eddy simulation of air entrainment under plunging breaking waves. *Coastal Engin.* **53** (8), 631–655.

MELVILLE, W. K. 1996 The role of surface-wave breaking in air-sea interaction. *Ann. Rev. Fluid Mech.* **28** (1), 279–321.

MELVILLE, W. K., VERON, F. & WHITE, C. J. 2002 The velocity filed under breaking waves: coherent structures and turbulence. *J. Fluid Mech.* **454**, 203–233.

MILLER, RL 1972 Study of air entrainment in breaking waves. *Eos T Am Geophys Un* **53** (4), 426–.

MILLER, S. J., SHEMDIN, O. H. & LONGUET-HIGGINS, M. S. 1991 Laboratory measurements of modulation of short-wave slopes by long surface-waves. *J. Fluid Mech.* **233**, 389–404.

MOSTERT, W., POPINET, S. & DEIKE, L. 2022 High-resolution direct simulation of deep water breaking waves: transition to turbulence, bubbles and droplets production. *J. Fluid Mech.* **942**.

MYRHAUG, D. & KJELDSEN, S. P. 1979 Breaking waves in deep water and resulting wave forces. In *Breaking waves in deep water and resulting wave forces*, pp. 2515–22. New York: Am. Inst. Min. Metall. Petrol. Eng.

OCHI, M. K. & TSAI, C. H. 1983 Prediction of occurrence of breaking waves in deep-water. *J. Phys. Ocean.* **13** (11), 2008–2019.

PERLIN, MARC, CHOI, WOOYOUNG & TIAN, ZHIGANG 2013 Breaking waves in deep and intermediate waters. *Ann. Rev. Fluid Mech.* **45**, 115–145.

PERLIN, M, HE, JH & BERNAL, LP 1996 An experimental study of deep water plunging breakers. *Phys Fluids* **8** (9), 2365–2374.

PHILLIPS, O. M. 1981 The dispersion of short wavelets in the presence of a dominant long-wave. *J. Fluid Mech.* **107**, 465–485.

PIZZO, N. E., DEIKE, LUC & MELVILLE, W. KENDALL 2016 Current generation by deep-water breaking waves. *J. Fluid Mech.* **803**, 275–291.

RAMBER, S. E. & GRIFFIN, O. M. 1987 A laboratory study of steep and breaking deep water waves. *J. Waterway, Port, Coastal, Ocean Eng.* **113** (5), 493–506.

RAPP, R. J. & MELVILLE, W. K. 1990 Laboratory measurements of deep-water breaking waves. *Philosophical Transactions of the Royal Society of London A: Mathematical, Physical and Engineering Sciences* **331** (1622), 735–800.

REYNOLDS, A. J. & TUCKER, H. J. 1975 Distortion of turbulence by general uniform irrotational strain. *J. Fluid Mech.* **68**, 673–693.

ROMERO, LEONEL, MELVILLE, W. KENDALL & KLESS, JESSICA M. 2012 Spectral energy dissipation due to surface wave breaking. *J. of Phys. Ocean.* **42** (9), 1421–1444.

SCHWARTZ, L. W. 1974 Computer extension and analytic continuation of stokes expansion for gravity-waves. *J. Fluid Mech.* **62** (FEB11), 553–578.

TIAN, ZHIGANG, PERLIN, MARC & CHOI, WOOYOUNG 2012 An eddy viscosity model for two-dimensional breaking waves and its validation with laboratory experiments. *Phys. Fluids* **24**, 036601.

TIAN, ZHIGAN, PERLIN, MARC & CHOI, WOOYOUNG 2018 Evaluation of a deep-water wave breaking criterion. *Phys. Fluids* **20**.

TUCKER, H. J. & REYNOLDS, A. J. 1968 Distortion of turbulence by irrotational plane strain. *J. Fluid Mech.* **32** (4), 657+.

VERON, FABRICE 2015 Ocean spray. *Ann. Rev. Fluid Mech.* **47** (1), 507–538.

WANG, A., IKEDA-GILBERT, C.M., DUNCAN, J. H., LATHROP, D. P., COOKER, M. J. & FULLERTON, A. M. 2018 The impact of a deep-water plunging breaker on a wall with its bottom edge close to the mean water surface. *J. Fluid Mech.* **843**, 680–721.

WANG, ZHAOYUAN, YANG, JIANMING & STERN, FREDERICK 2016 High-fidelity simulations of bubble, droplet and spray formation in breaking waves. *J. Fluid Mech.* **792**, 307–327.

WATANABE, Y & SAEKI, H 2002 Velocity field after wave breaking. *Int. J. Num. Meth. Fluids* **39** (7), 607–637.

WATANABE, Y, SAEKI, H & HOSKING, RJ 2005 Three-dimensional vortex structures under breaking waves. *J Fluid Mech* **545**, 291–328.

ZHONG, XIAOXU & LIAO, SHIJUN 2018 On the limiting stokes wave of extreme height in arbitrary water depth. *J. Fluid Mech.* **843**, 653–679.

## Appendix A.  Tables of profile alignment data and polynomial coefficients for crest point and jet tip trajectories

Table 5 shows standard deviation of the $x$ and $y$ location of the wave crest point at the time of jet formation, which is discussed in § 3.1. The wave profile alignment in $x$, $y$, and $t$ for each of the 10 runs for the weak, moderate, and strong plunging breakers is shown in table 6 and discussed in § 3.1. Finally, table 7 shows the coefficients and $R^2$ values from the fitting of the wave crest and jet tip trajectories, which are discussed in § 3.2.

| | Pre-alignment | | | |
|---|---|---|---|---|
| Breaker Type | $\sqrt{\langle[x_c^f - \langle x_c^f\rangle]^2\rangle}$ | $\sqrt{\langle[y_c^f - \langle y_c^f\rangle]^2\rangle}$ | $\sqrt{\langle[x_c^i - \langle x_c^i\rangle]^2\rangle}$ | $\sqrt{\langle[y_c^i - \langle y_c^i\rangle]^2\rangle}$ |
| Weak | 4.93 | 0.46 | 5.25 | 0.35 |
| Moderate | 6.13 | 0.34 | 5.94 | 0.28 |
| Strong | 7.97 | 0.31 | 7.08 | 0.43 |

| | Post-alignment | | | |
|---|---|---|---|---|
| Breaker Type | $\sqrt{\langle[x_c^f - \langle x_c^f\rangle]^2\rangle}$ | $\sqrt{\langle[y_c^f - \langle y_c^f\rangle]^2\rangle}$ | $\sqrt{\langle[x_c^i - \langle x_c^i\rangle]^2\rangle}$ | $\sqrt{\langle[y_c^i - \langle y_c^i\rangle]^2\rangle}$ |
| Weak | 1.27 | 0.21 | 1.77 | 0 |
| Moderate | 1.54 | 0.44 | 1.49 | 0 |
| Strong | 1.97 | 0.47 | 1.39 | 0 |

Table 5: The standard deviation in mm of the position of the wave crest point at the time of jet formation, $x_c^f$ and $y_c^f$, and jet impact, $x_c^i$ and $y_c^i$, before and after breaker profile alignment.

| | Weak | | | | Moderate | | | | Strong | | | |
|---|---|---|---|---|---|---|---|---|---|---|---|---|
| Run | $\Delta x_b^i$ | $\Delta y_c^i$ | $\Delta t^i$ | $\Delta t^f$ | $\Delta x_b^i$ | $\Delta y_c^i$ | $\Delta t^i$ | $\Delta t^f$ | $\Delta x_b^i$ | $\Delta y_c^i$ | $\Delta t^i$ | $\Delta t^f$ |
| 1 | 1.3 | 0.4 | -2 | 0.8 | 5.3 | 0.2 | -2.9 | -3.8 | 14 | -0.3 | -7.1 | -3.5 |
| 2 | -6.9 | 0.5 | 4.2 | 0.8 | -3.3 | 0.4 | 3.2 | 2.3 | 3.6 | -0.4 | -2.5 | -0.5 |
| 3 | -3.8 | 0.3 | 1.1 | -0.8 | 2.5 | 0.2 | -1.4 | -0.8 | -0.2 | -0.2 | -0.9 | -0.5 |
| 4 | -3.2 | 0.1 | 1.1 | -0.8 | 7.0 | 0.2 | -6.0 | -3.8 | -0.5 | -0.4 | 2.2 | 1.1 |
| 5 | -1.5 | -0.2 | 1.1 | -0.8 | 4.3 | 0.2 | -2.9 | 2.3 | 1.6 | -0.6 | 2.2 | 2.6 |
| 6 | 2.3 | -0.2 | -3.5 | -5.4 | 7.9 | -0.2 | -4.5 | -2.3 | 2.2 | 0.4 | -0.9 | -3.5 |
| 7 | 4.6 | 0.0 | -3.5 | -3.8 | -1.2 | 0.0 | 0.2 | -0.8 | 4.1 | 0.6 | -5.5 | -9.7 |
| 8 | -3.8 | -0.1 | 4.2 | 8.5 | -5.2 | -0.3 | 1.7 | -0.8 | -14 | 0.3 | 6.8 | 7.2 |
| 9 | 1.1 | -0.3 | 2.6 | 3.8 | -8.4 | -0.3 | 6.3 | 3.8 | -7.7 | 0.5 | 3.7 | 4.2 |
| 10 | 9.9 | -0.6 | -5.1 | -2.3 | -9.1 | -0.5 | 6.3 | 3.8 | -3.3 | 0.0 | 2.2 | 2.6 |

Table 6: Table of the variations in positions and times of the crest point (denoted by subscript $c$) and the back face (denoted by subscript $bf$) at the times of jet formation (denoted by superscript $f$) and jet impact (denoted by superscript $i$) where, $\Delta x_b^i = x_b^i - \langle x_b^i\rangle$, $\Delta y_c^i = y_c^i - \langle y_c^i\rangle$, $\Delta t^i = t^i - \langle t^i\rangle$ and $\Delta t^f = t^f - \langle t^f\rangle$ for 10 realizations of each breaker. Times are given in ms and distances are given in mm. The distances and times in this table can be nondimensionalized by $\langle r_x^i\rangle$, $\langle r_y^i\rangle$, and $\langle\Delta t^{f\text{-}i}\rangle$ given in table 2.

| $f(x)$ | $a$ | $b$ | $c$ | $d$ | $R^2$ |
|--------|-----|-----|-----|-----|-------|
| Weak | | | | | |
| $\langle \tilde{x}_c(t) \rangle$ | 4.33e-01 | 3.87e-02 | 1.31e-01 | -3.05e-05 | 0.999 |
| $\langle \tilde{y}_c(t) \rangle$ | 1.99e-01 | 7.73e-03 | -3.12e-03 | -5.49e-06 | 0.860 |
| $\langle \tilde{x}_j(t) \rangle$ | -1.70e-01 | 2.09e-02 | 1.69e-01 | 5.53e-03 | 0.999 |
| $\langle \tilde{y}_j(t) \rangle$ | -8.82e-01 | -5.30e-01 | -8.78e-02 | -5.21e-03 | 0.999 |
| Moderate | | | | | |
| $\langle \tilde{x}_c(t) \rangle$ | 3.23e-01 | 3.31e-02 | 1.31e-01 | -1.35e-05 | 0.999 |
| $\langle \tilde{y}_c(t) \rangle$ | 1.85e-01 | 5.66e-03 | -1.96e-03 | -1.26e-06 | 0.916 |
| $\langle \tilde{x}_j(t) \rangle$ | -2.32e-01 | 1.18e-02 | 1.75e-01 | 7.15e-03 | 0.999 |
| $\langle \tilde{y}_j(t) \rangle$ | -1.01e-01 | -4.18e-01 | -9.24e-02 | -6.43e-03 | 0.999 |
| Strong | | | | | |
| $\langle \tilde{x}_c(t) \rangle$ | 2.92e-01 | 2.84e-02 | 1.32e-01 | 1.18e-05 | 0.999 |
| $\langle \tilde{y}_c(t) \rangle$ | 1.45e-01 | -6.75e-03 | -2.15e-03 | -2.29e-06 | 0.921 |
| $\langle \tilde{x}_j(t) \rangle$ | -3.44e-01 | -4.14e-03 | 1.77e-01 | 7.95e-03 | 0.999 |
| $\langle \tilde{y}_j(t) \rangle$ | -4.56e-01 | -4.58e-01 | -9.55e-02 | -6.92e-03 | 0.999 |

Table 7: A table of the coefficients from fitting third-order polynomials to the ensemble averaged wave crest point trajectories, $\langle \tilde{x}_c(t) \rangle$ and $\langle \tilde{y}_c(t) \rangle$, and the jet tip trajectories, $\langle \tilde{x}_j(t) \rangle$ and $\langle \tilde{y}_j(t) \rangle$, for the weak, moderate, and strong breakers. The polynomials have the form $x(t) = at^3 + bt^2 + ct + d$ where $x$ is the position in meters, $t$ is time in seconds, $a - d$ are the polynomial coefficients, and $R^2$ is the coefficient of determination from the fitting procedure. The polynomials were fitted to data recorded in the time range $t^f - 13.8 \ ms \leqslant t \leqslant t^i + 13.8 \ ms$.